\documentclass[prl,aps,superscriptaddress,footinbib,twocolumn]{revtex4-2}%
\usepackage[latin9]{inputenc}
\usepackage{color}
\usepackage{amsmath}
\usepackage{amssymb}
\usepackage{stmaryrd}
\usepackage{graphicx}
\usepackage[unicode=true,
bookmarks=true,bookmarksnumbered=false,bookmarksopen=false,
breaklinks=false,pdfborder={0 0 1},backref=false,colorlinks=true]
{hyperref}
\hypersetup{
	linkcolor=magenta, urlcolor=blue, citecolor=blue, pdfstartview={FitH}, hyperfootnotes=false, unicode=true}

\makeatletter

\usepackage{amsfonts}\usepackage{tabularx}\usepackage{dcolumn}\usepackage{bm}\usepackage{graphicx}\usepackage{epstopdf}
\usepackage{times}
\hypersetup{urlcolor=blue}

\makeatother

\begin{document}
		\title{Digital simulation of projective non-Abelian anyons with 68 superconducting qubits}
		
		\affiliation{School of Physics, ZJU-Hangzhou Global Scientific and Technological Innovation Center, Interdisciplinary Center for Quantum Information, and Zhejiang Province Key Laboratory of Quantum Technology and Device, Zhejiang University, Hangzhou 310027, China\\
			$^2$ Center for Quantum Information, IIIS, Tsinghua University, Beijing 100084, China\\
			$^3$ Theoretical Physics Division, Chern Institute of Mathematics and LPMC, Nankai University, Tianjin 300071, China\\
			{$^{4}$ Hefei National Laboratory, Hefei 230088, China}\\
			$^{5}$ Shanghai Qi Zhi Institute, 41th Floor, AI Tower, No. 701 Yunjin Road, Xuhui District, Shanghai 200232, China}
		\author{Shibo Xu$^{1}$}\thanks{These authors contributed equally to this work.}
		\author{Zheng-Zhi Sun$^{2}$}\thanks{These authors contributed equally to this work.}
		\author{Ke Wang$^{1}$}\thanks{These authors contributed equally to this work.}
		\author{Liang Xiang$^{1}$}
		\author{Zehang Bao$^{1}$}
		\author{Zitian Zhu$^{1}$}
		\author{Fanhao Shen$^{1}$}
		\author{Zixuan Song$^{1}$}
		\author{Pengfei Zhang$^{1}$}
		\author{Wenhui Ren$^{1}$}
		\author{Xu Zhang$^{1}$}
		\author{Hang Dong$^{1}$}
		\author{Jinfeng Deng$^{1}$}
		\author{Jiachen Chen$^{1}$}
		\author{Yaozu Wu$^{1}$}
		\author{Ziqi Tan$^{1}$}
		\author{Yu Gao$^{1}$}
		\author{Feitong Jin$^{1}$}
		\author{Xuhao Zhu$^{1}$}
		\author{Chuanyu Zhang$^{1}$}
		\author{Ning Wang$^{1}$}
		\author{Yiren Zou$^{1}$}
		\author{Jiarun Zhong$^{1}$}
		\author{Aosai Zhang$^{1}$}
		\author{Weikang Li$^{2}$}
		\author{Wenjie Jiang$^{2}$}
		\author{Li-Wei Yu$^{3}$}
		\author{Yunyan Yao$^{1}$}
		\author{Zhen Wang$^{1, 4}$}
		\author{Hekang Li$^{1}$}
		\author{Qiujiang Guo$^{1, 4}$}
		\author{Chao Song$^{1, 4}$}\email{chaosong@zju.edu.cn}
		\author{H. Wang$^{1, 4}$}\email{hhwang@zju.edu.cn}
		\author{Dong-Ling Deng$^{2, 4, 5}$}\email{dldeng@tsinghua.edu.cn}

		\begin{abstract}
			\textbf{Non-Abelian anyons are exotic quasiparticle excitations hosted by certain topological phases of matter. They break the fermion-boson dichotomy and obey non-Abelian braiding statistics: their interchanges yield unitary operations, rather than merely a phase factor, in a space spanned by topologically degenerate wavefunctions. They are the building blocks of topological quantum computing. However, { experimental observation of non-Abelian anyons and their characterizing braiding statistics is notoriously challenging and has remained elusive hitherto}, in spite of various theoretical proposals. Here, we report an experimental quantum digital simulation of projective non-Abelian anyons and their braiding statistics with up to 68 programmable superconducting qubits arranged on a two-dimensional lattice. By implementing the ground states of the toric-code model  with twists  through quantum circuits,  we demonstrate that twists exchange electric and magnetic charges and behave as a particular type of non-Abelian anyons, i.e., the Ising anyons. In particular,  we show experimentally that these twists follow the fusion rules and non-Abelian braiding statistics of the Ising type, and can be explored to encode topological logical qubits. Furthermore,  we demonstrate how to implement both single- and two-qubit logic gates through applying a sequence of elementary Pauli gates on the underlying physical qubits. %
				Our results demonstrate a versatile quantum digital approach for simulating non-Abelian anyons, offering a new lens into the study of such peculiar quasiparticles.  
			}
		\end{abstract}
		
		\maketitle
		
		Quantum theory classifies all fundamental particles in nature as either bosons or fermions \cite{Griffiths2018Introduction}. For instance, photons are bosons and electrons are fermions. This dichotomy classification has profound implications and plays a crucial  role in understanding a variety of physical phenomena, ranging from metal-insulator transitions \cite{Imada1998Metal} to superconductivity \cite{Tinkham2004Introduction} and Bose-Einstein condensation \cite{Dalfovo1999Theory}. However, in two dimensions it is possible that emergent particles (quasiparticles) would circumvent this dichotomy principle and obey anyonic statistics \cite{Wilczek1982Quantum}, where their exchange of positions would result in a generic phase factor that is neither $0$ nor $\pi$ (as for bosons or fermions), or even  a unitary operation that shifts the system between different topologically degenerate states \cite{nayak2008non}. These quasiparticles are dubbed anyons \cite{Wilczek1990Fractional}. 
		
		While braiding Abelian anyons only leads to a phase factor, braiding non-Abelian anyons would lead to a unitary transformation \cite{nayak2008non,stern2010non,Wilczek1990Fractional,moore1991nonabelions,Wen1991NonAbelian}. This prominent property gives rise to the notion of topological quantum computation \cite{Kitaev2003FaultT,Freedman2002Modular, nayak2008non,Sarma2006Topological,Stern2013Topological}, where quantum information is encoded nonlocally and quantum computations are implemented by braiding and fusing non-Abelian anyons. The nonlocal encoding and the topological nature of braiding make topological quantum computation naturally immune to local errors, thus providing intrinsic fault tolerance at the level of hardware. However, despite numerous theoretical proposals for realizing non-Abelian anyons with a wide range of systems \cite{kitaev2006anyons,Barkeshli2013Twist,Teo2014Unconventional,Bombin2010Topological,Zheng2015Demonstrating,Brown2017Poking,Alicea2011non,ivanov2001non,Bonderson2006Detecting, clarke2013exotic,Tantivasadakarn2022Shortest,Liu2022Methods, Kalinowski2022NonAbelian}, including fractional quantum Hall states \cite{moore1991nonabelions,Bonderson2006Detecting}, cold atoms \cite{Deng2015Proposal}, topological superconductors \cite{ivanov2001non}, and Majorana zero modes \cite{Alicea2011non}, the direct experimental observation of non-Abelian anyons and their braiding statistics  still remains elusive so far \cite{Banerjee2018Observation,Kasahara2018Majorana,Dolev2008Observation,Bartolomei2020Fractional}.  Meanwhile, recent developments with quantum processors have demonstrated vast potential in simulating exotic phases of matter \cite{Satzinger2021Realizing,Dumitrescu2022Dynamical,Kyprianidis2021Observation,Mi2022Timecrystallinea,Zhang2022Digital} and demonstrating  quantum error-correcting codes~\cite{google_surface_code,ustc_surface_code_2022, eth_surface_code, delft_surface_code},  giving rise to intriguing opportunities for simulating and exploring non-Abelian anyons with these highly controllable systems \cite{Andersen2022Observation}.

		\begin{figure*}[htb]
			\includegraphics[width=1\linewidth]{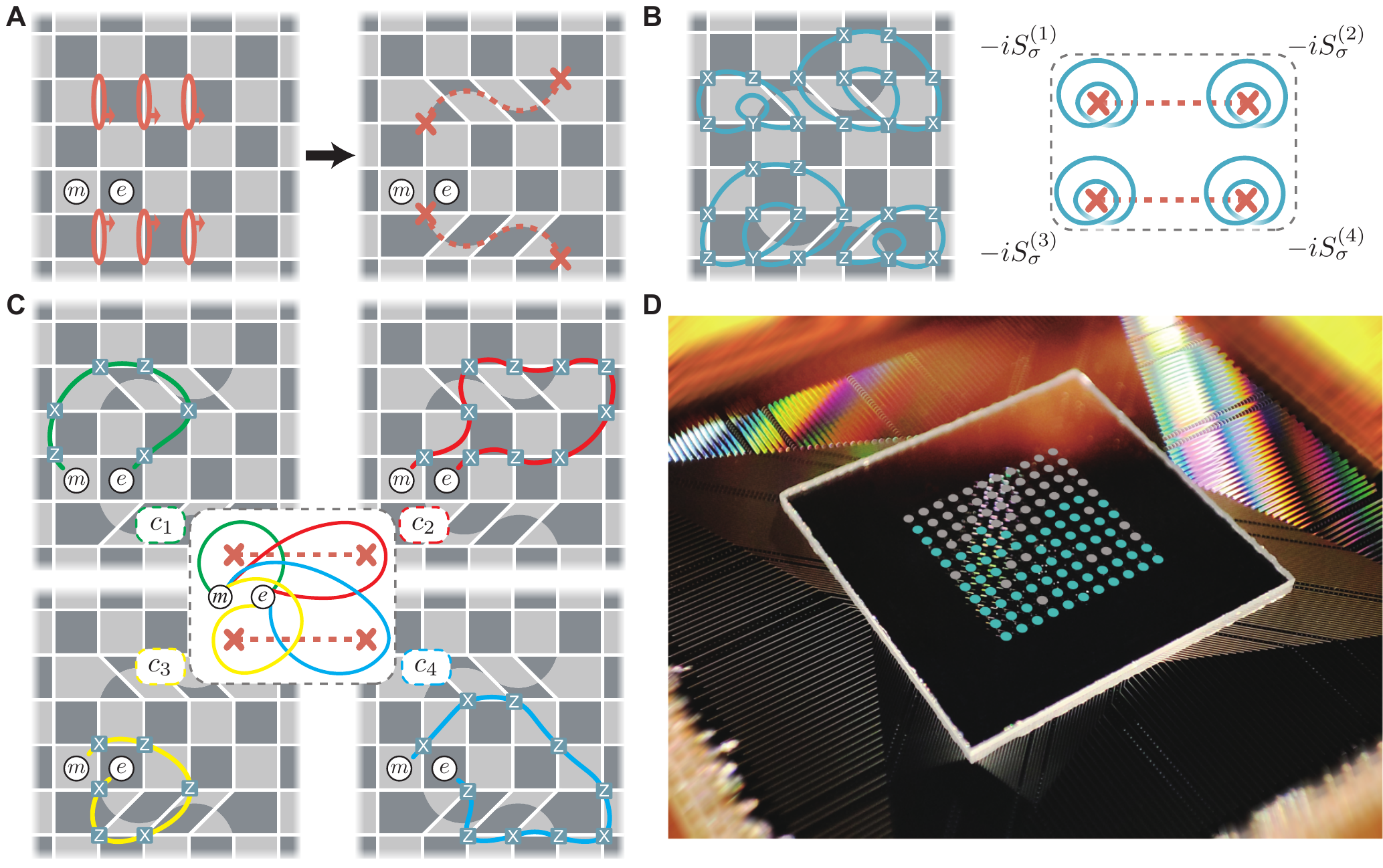}
			\caption{ \textbf{ Simulating non-Abelian anyons with twists and the photograph of the experimental quantum processor I.} (\textbf{A})  The deformation of the toric-code model with two pairs of twists, which behave as Ising anyons under braiding and fusion. Two Abelian topological charges $e$ and $m$ are illustrated, with $e$ anyons living on the dark plaquettes and $m$ anyons on the light ones. The twists are formed by rearranging the circled bonds and marked with the cross symbol ``$\times$". (\textbf{B})  {The string operators $-i{{S}_\sigma }$ that specify the generalized charge of the non-Abelian twists. The fusion rules between the twists and other charges can be demonstrated by using these operators.} In the right panel, we use a conceptual figure to sketch the relative positions of the twists and the corresponding string operators, omitting the details of the lattice sites and the constitutive local Pauli operators for simplicity and better illustration. 
				(\textbf{C})  {Four Majorana operators are used to describe the four non-Abelian twists and the corresponding conceptual figure. The fusion and braiding of non-Abelian twists can be conveniently implemented with these operators \cite{Bombin2010Topological}.} (\textbf{D})  A photograph of the superconducting quantum processor I with the chosen 68 qubits used in our experiment highlighted in cyan. }\label{fig-sketch-deform-68}
		\end{figure*}
		
		{Here we report an experimental quantum digital simulation of projective non-Abelian anyons and their nontrivial braiding statistics with two superconducting processors (labeled as I and II), which are designed to carry arrays of 11$\times$11 and 6$\times$6
			frequency-tunable transmon qubits respectively. We select  68 (30) qubits on processor I (II) featuring a median lifetime  of $109.8$ $\mu$s ($139.8$ $\mu$s), and median fidelities of simultaneous single- and two-qubit gates above $99.91\%$ and $99.4\%$ ($99.95\%$ and $99.5\%$) respectively, for carrying out our experiments. We prepare the twisted toric-code ground states through efficient quantum circuits with a depth up to 43. We demonstrate that the twists exchange electric and magnetic charges when an odd number of them are winded around, and their braiding and fusion rules resemble that of Ising-type anyons. In addition, we show that these twists can be explored to encode topological logical qubits and both single- and two-qubit logic gates can be implemented by applying a sequence of elementary Pauli gates on the physical superconducting qubits. Based on this, we are able to prepare a logical two-qubit Bell state on the processor I and a logical three-qubit Greenberger-Horne-Zeilinger (GHZ) state on the processor II, showing the existence of multipartite entanglement among the logical qubits. %
		}
		
		\vspace{.5cm}
		
		\noindent\textbf{\large{}Framework and experimental setups}	
		
		\noindent  We consider the toric-code model with qubits living on the vertexes of a square lattice described by the following Hamiltonian \cite{Kitaev2003FaultT,Wen2003QuantumPRL}: 
		\begin{align}\label{eq-Hamiltonian-surface}
			{H} =  - \sum\limits_{\bf k} {{{\bf{A}}_{\bf k}}}, \; {{{\bf{A}}_{\bf k}}}={{X}_{\bf k}}{{Z}_{\bf {k} + {i}}}{{Z}_{\bf {k} + {j}}}{{X}_{\bf {k} + {i} + {j}}}.
		\end{align}
		Here ${\bf k} = \left( {a,b} \right)$ indexes the spins in the $a$-th row and $b$-th column and ${\bf i} = \left( {1,0} \right)$, ${\bf j} = \left( {0,1} \right)$. $X$, $Y$, and $Z$ are Pauli operators. Noting that all the plaquette operators ${\bf{A}}_{\bf k}$ commute with each other, the ground state of this Hamiltonian can be simply described by the condition $\langle\textbf{A}_{\bf k}\rangle=1$ for all ${\bf k}$. There are two types of quasiparticle excitations (corresponding to $\langle\mathbf{A}_{\mathbf {k}}\rangle=-1$), dubbed $e$ anyons (or ``electric charges")  living at the dark plaquettes and $m$ anyons (or ``magnetic charges") living at the light plaquettes, respectively (see Fig. \ref{fig-sketch-deform-68}\textbf{A}). These are Abelian anyons \cite{Kitaev2003FaultT,Wen2003QuantumPRL}, and they can be created and moved by string operators, which are products of sequences of Pauli operators, as shown in Fig. \ref{fig-sketch-deform-68}\textbf{B}. When an $e$ anyon is fused with an $m$ anyon, one obtains an $\epsilon$ particle that behaves as a fermion. { For our purpose of simulating non-Abelian anyons in a digital fashion, we consider introducing dislocations in the lattice \cite{Bombin2010Topological,Kitaev2012Models}.} For instance, in Fig. \ref{fig-sketch-deform-68}\textbf{A} we deform the lattice so as to obtain two pairs of pentagonal plaquettes, and modify the Hamiltonian with new pentagonal plaquette operators accordingly. After the deformation, each pentagonal plaquette hosts a twist, which involves a lattice site that is shared by three (rather than four) neighboring plaquettes. As shown in Fig. \ref{fig-sketch-deform-68}\textbf{C}, due to the dislocation a string  winding around a twist cannot close, and consequently an $e$ anyon winds around a twist will become an $m$ anyon or vice versa. Theoretically, { it has been predicted that  twists resemble non-Abelian Ising anyons when braided and fused \cite{Bombin2010Topological}.} %

		\begin{figure*}[htb]
			\includegraphics[width=1\linewidth]{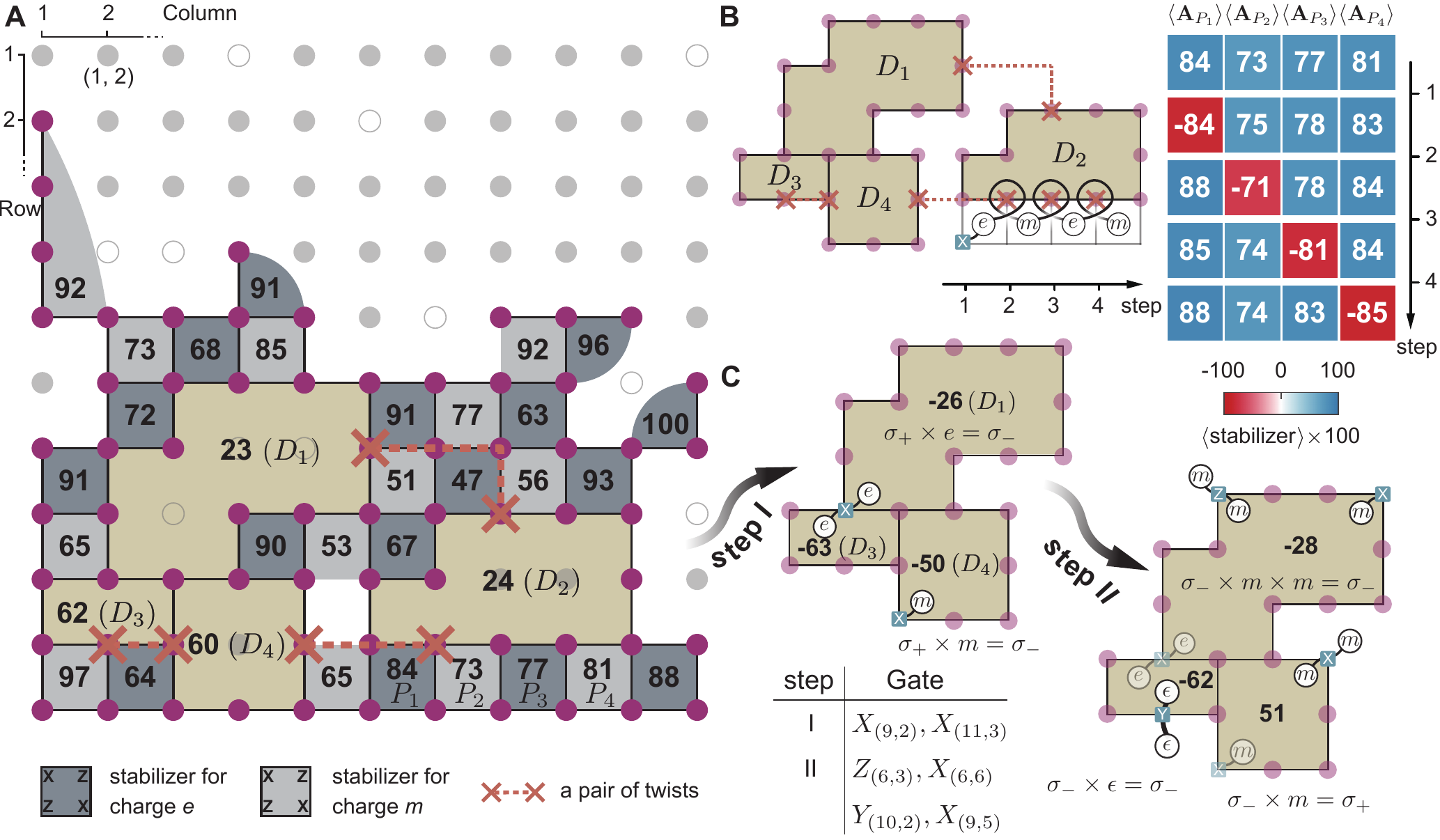}
			\caption{\label{GroundStateAndFusion} %
				\textbf{The processor I and demonstration of the fusion rules}. (\textbf{A})  Layout of the  processor and the measured stabilizer values of the deformed toric-code ground state.  
				The purple dots represent the chosen qubits used in our experiment, whereas the grey (white) dots denote the functional but unused (non-functional) qubits. The black line connecting two adjacent qubits shows where a controlled-Z (CZ) gate can be implemented. The dark (light) grey plaquettes host $e$ ($m$) anyons, and the brown blocks denote the natural defects explored in our experiment to implement twists.  The stabilizer for the smallest square reads ${{\bf{A}}_{\bf k}} = {{X}_{\bf k}}{{Z}_{\bf k + {i}}}{{Z}_{\bf k + {j}}}{{X}_{\bf k + {i} + {j}}}$, and that for each defect is the product of all ${\bf{A}_k}$ it encloses. { The three pairs of crosses connected by dashed lines denote the six twists used to simulate non-Abelian anyons.} The integer number in each plaquette shows the corresponding measured stabilizer value (in percentage) of the prepared deformed toric-code state. 
				(\textbf{B})  The sketch of the defects with eight twists. In the experiment, we create {an $e$ anyon by applying the gate $X_{(11,6)}$ and then move it} to wind around the three marked twists below $D_2$ successively. We measure the four stabilizers 
				($\mathbf{A}_{P_1}$, $\mathbf{A}_{P_2}$, $\mathbf{A}_{P_3}$, and $\mathbf{A}_{P_4}$) at each step and their corresponding values are shown in the right panel.  (\textbf{C})  Demonstration of the fundamental fusion rules. We initialize the system to be a ground state with the generalized charges being $i$ for the chosen six twists marked in (\textbf{A}) ($S_\sigma=i$, corresponding to six $\sigma_+$ twists, see \cite{Bombin2010Topological} and {supplementary materials I.D}). {We then create $e$, $m$, and $\epsilon$ quasiparticles,} and measure the stabilizer values corresponding to the defects, which specify the fusion results. The integer numbers in the defects show the measured stabilizer values (in percentage), respectively. In the lower left panel, the table indicates the Pauli operations used to generate the quasiparticles. }
		\end{figure*}

		{Our experiments are performed on two flip-chip superconducting quantum processors  I and II.} Processor I (II) encapsulates $11\times11$ ($6\times6$) frequency-tunable transmon qubits arranged in a square lattice, with tunable couplers connecting adjacent qubits. Each qubit capacitively couples to its own readout resonator for qubit state measurements. { We select 68 (30) qubits on processor I (II) to simulate non-Abelian anyons and their associated braiding statistics. } Through optimizing device fabrication and controlling process, we push the median lifetime of the qubits on processor I (II) to $109.8$ $\mu$s ({ $139.8$ $\mu$s}) and the median simultaneous single- and two-qubit gate fidelities greater than $99.91\%$ and $99.4\%$ ({ $99.95\%$ and $99.5\%$}), respectively. { The chosen  30 qubits on the processor II} form a regular rectangular lattice and the desired twists are artificially created to some extent. {  Whereas, the $68$ qubits on the processor I are more irregularly distributed} due to limited capacities in both wirings of our dilution refrigerator and measurement electronics, and twists can be constructed by taking advantages of the imperfect geometry. In the main text, we mainly discuss the results obtained from  processor I (Fig. \ref{fig-sketch-deform-68}\textbf{D}), so as to stress that our approach for simulating non-Abelian anyons bears the merit of generally applicable to quantum processors with unintended imperfections. {The experiments carried out on  processor II yield similar results, which we present in the supplementary materials for completeness and comparison.}

		\begin{figure*}[htb]
			\includegraphics[width=1\linewidth]{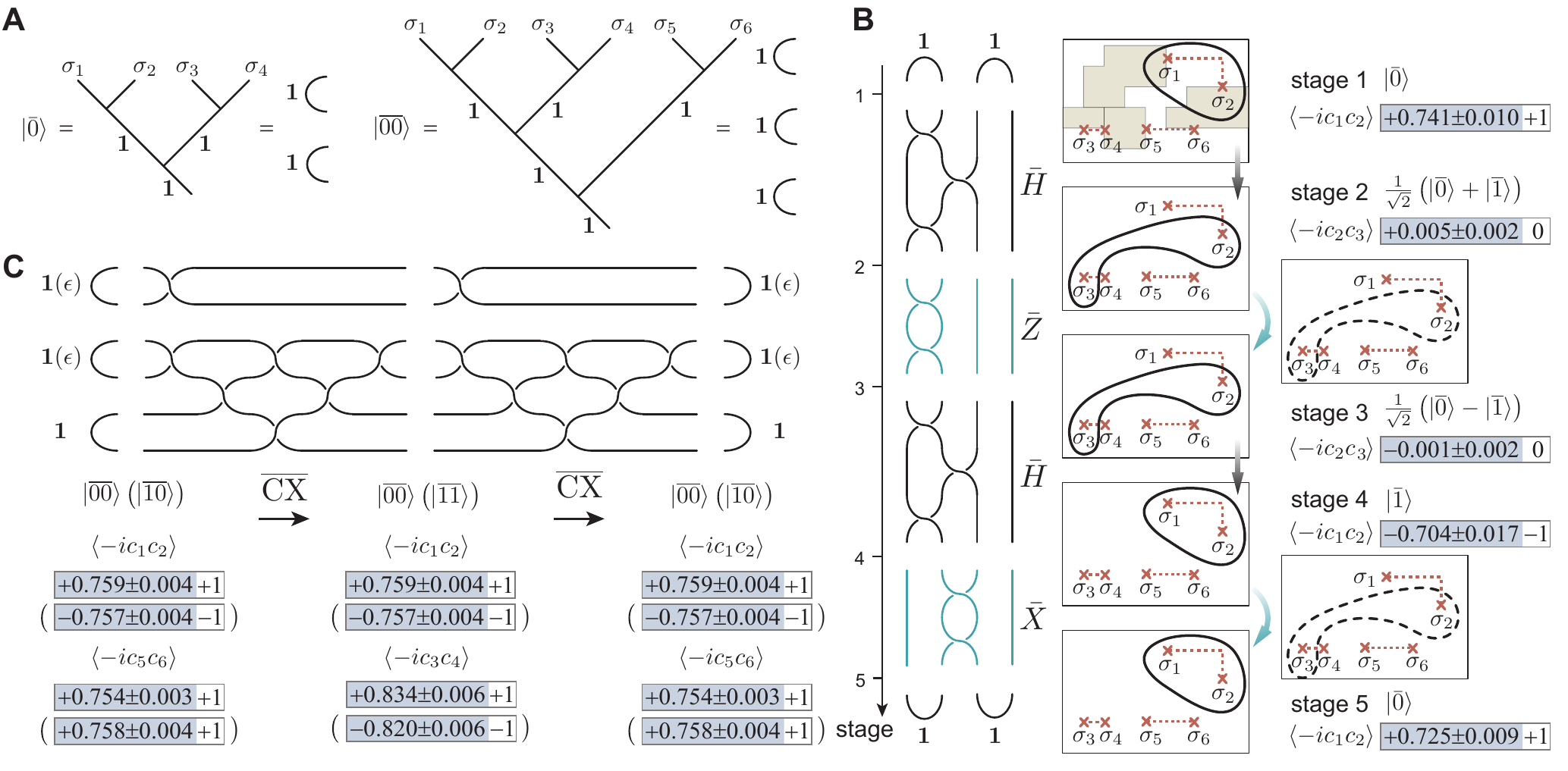}
			\caption{ %
				\textbf{ Encoding of logical qubits with twists and demonstration of logic gates.} (\textbf{A})  The encoding scheme described with fusion trees \cite{nayak2008non}. Here, for succinctness we only show part of the encoded logical bases $|\bar{0}\rangle$ and $|\overline{00}\rangle$ (supplementary materials I.F). 
				{(\textbf{B})  On the left shows the braiding sequence corresponding to the process of ${{\bar X\bar H\bar Z\bar H}}\left| {\bar 0} \right\rangle  = \left| {\bar 0} \right\rangle $, with $\bar H$ realized by Majorana tracking (black lines) and other operations realized by implementing the corresponding string operators on physical qubits (blue lines). 
					Time flows from up to down. 
					The top middle panel sketches the defects and the three pairs of twists (denoted by $\sigma_i$, $i=1,2,3,4,5,6$), with the black solid circle denoting the string operator of the logical $\bar{Z}$ observable (corresponding to the Majorana correlator $-ic_1c_2$) for the first logical qubit. 
					During the braiding procedure, the correspondence between the logical $\bar{Z}$ observable and Majorana correlator may change according to Majorana tracking, as outlined in the lower four panels in the middle column.
					The right column shows the logical states at each stage, which are verified by measuring the Majorana correlators corresponding to logical $\bar{Z}$ observable, with the experimental (shaded values with error bars) and ideal (unshaded integers) results following. The data are presented as mean values $\pm$ standard error of the mean with a sample size of five. The two black dashed circles sketch the string operators applied to the physical qubits, which correspond to  the logic $\bar{Z}$ and $\bar{X}$ gates, respectively.
				}
				(\textbf{C})  The braiding sequence corresponding to applying two logic controlled-X ($\overline{\text{CX}}$) gates consecutively on the state $|\overline{00}\rangle$ or $|\overline{10}\rangle$. Time flows from left to right. We measure the corresponding Majorana correlators at each step and show their measured values in the lower panel. {See {supplementary materials I.F} for details. }
			}\label{fig-quantum-gates-68}
		\end{figure*}

		\vspace{.5cm}
		\noindent\textbf{\large{}Topological charges and fusion rules}{\large\par}
		
		\noindent { We first demonstrate the  simulation of both Abelian and non-Abelian anyons and the fundamental fusion rules.} Based on the geometric structure for the chosen 68 qubits on processor I and the available connections between them, we use a quantum circuit with a circuit depth of 43 (containing $294$ single-qubit rotations and $113$ two-qubit controlled-Z gates, see {fig. S12})
		to prepare the ground state of the corresponding stabilizer Hamiltonian ({supplementary materials I.G}). In Fig. \ref{GroundStateAndFusion}\textbf{A}, we sketch the geometric structure of the chosen $68$ qubits and  plot the  individual stabilizer values measured after the ground state preparation. We achieve an average stabilizer value of $0.73$%
		, which is notable given the fact that certain stabilizers involve multi-qubit (up to $14$ qubits) measurements.  The deviation between the ideal theoretical prediction and experimental result is mainly attributed to limited gate fidelity and coherence time. All stabilizer values are positive, implying that there are no anyon excitations for the ground states. { See also {fig. S14}
			for experimental results from processor II, where an average stabilizer value of  $0.86$ is obtained.}

		With twists, the system can hosts six generalized topological charges \cite{Bombin2010Topological}: $\mathbf{1}$, $e$, $m$, $\epsilon$, $\sigma_+$, and $\sigma_-$. They obey the following fundamental fusion rules that can be verified using proper string operators
		\begin{equation}\label{Eq:Fusionrules}
			\begin{aligned}
				& \sigma_{\pm} \times e =\sigma_{\pm} \times m=\sigma_{\mp}, \quad \sigma_{\pm} \times \epsilon=\sigma_{\pm},  \\
				&	\sigma_{\pm} \times \sigma_{\pm} =1+\epsilon, \quad \sigma_{\pm} \times \sigma_{\mp}=e+m.
			\end{aligned}
		\end{equation}		
		In our experiments, we demonstrate only some of the fusion rules for simplicity and concreteness. The remaining fusion rules either follow trivially or can be demonstrated in a similar way. After preparing the ground state of the Hamiltonian with twists, { we create an $e$ anyon and move it to wind around the twists located at the bottom of the $D_2$ defect} (Fig. \ref{GroundStateAndFusion}\textbf{B}), through applying corresponding string operators on the relevant qubits.  We measure the four stabilizers (labeled by $\mathbf{A}_{P_1}$, $\mathbf{A}_{P_2}$, $\mathbf{A}_{P_3}$, and $\mathbf{A}_{P_4}$, respectively) below $D_2$  at each step of this process and plot their values in the right panel of Fig. \ref{GroundStateAndFusion}\textbf{B}. It is clear that, at the beginning, all measured stabilizer values are positive, indicating that there is no anyon excitation in the system. { After the creation of the $e$ anyon,  $\langle \mathbf{A}_{P_1}\rangle$ becomes negative, which indicates that there is an excitation at plaquette $P_1$ (an $e$ anyon).} This  $e$ anyon is then moved around the first twist and becomes an $m$ anyon, which is confirmed in the experiment by the observation that $\langle \mathbf{A}_{P_2}\rangle$ becomes negative whereas $\langle \mathbf{A}_{P_1}\rangle$ changes back to be positive at step 2. The $m$ anyon is further moved around the second twist and becomes an $e$ anyon, as confirmed by the measured stabilizer values at step 3. At step $4$, we further move the $e$ anyon around the third twist and it becomes an $m$ anyon again.  This clearly shows that winding around an odd number of twists exchanges electric and magnetic charges. See also {fig. S15}
		for experimental results from processor II, where the fact that an $e$ anyon winding around two twists will remain the same type is demonstrated as well.

		In Fig. \ref{GroundStateAndFusion}\textbf{C}, we plot the measured results for 	$\mathbf{A}_{D_1}$, $\mathbf{A}_{D_3}$, and $\mathbf{A}_{D_4}$ (in the experiment, we have in fact measured all the stabilizers. Here, for better illustration we only plot three of them that are most relevant for the discussion), which determine the topological charges for the relevant twists marked as crosses in  Fig. \ref{GroundStateAndFusion}\textbf{A} and hence verify the fusion rules. At the beginning, we prepare the system to the ground state of the stabilizer Hamiltonian, where all three measured stabilizer values are positive as shown  in Fig. \ref{GroundStateAndFusion}\textbf{A}. In other words, the system possesses six $\sigma_+$ twists at the beginning. { We then create a pair of $e$ anyons and an $m$ anyon by applying the $X$ gate on both qubits at sites $(9,2)$ and $(11,3)$ respectively,} as depicted in the left upper panel of Fig.  \ref{GroundStateAndFusion}\textbf{C}. We measure the stabilizers $\mathbf{A}_{D_1}$, $\mathbf{A}_{D_3}$, and $\mathbf{A}_{D_4}$, and find $\langle\mathbf{A}_{D_1}\rangle=-0.265\pm0.015$, %
		$\langle\mathbf{A}_{D_3}\rangle=-0.629\pm0.013$, and $\langle\mathbf{A}_{D_4}\rangle=-0.496\pm0.027$, consistent with two $\sigma_-$ twists at the same positions where the original $\sigma_+$ twists live. This verifies the fusion rules $\sigma_+\times e =\sigma_+\times m =\sigma_-$.

		{ We further create five  $m$ anyons and one pair of $\epsilon$ anyons, with their locations depicted in the right lower panel of Fig. \ref{GroundStateAndFusion}\textbf{C}.} We find that $\langle\mathbf{A}_{D_1}\rangle$ and  $\langle\mathbf{A}_{D_3}\rangle$ remain negative ($\langle\mathbf{A}_{D_1}\rangle=-0.281\pm0.012$ and $\langle\mathbf{A}_{D_3}\rangle=-0.620\pm0.005$), but $\langle\mathbf{A}_{D_4}\rangle=0.510\pm0.017$ becomes positive. This demonstrates the fusion rules of $\sigma_-\times m = \sigma_+$ and $\sigma_-\times \epsilon=\sigma_-$. 
		The demonstrated rule $\sigma_-\times \epsilon=\sigma_-$ implies that adding an $\epsilon$ quasiparticle to $\sigma_-$ will not change the total topological charge, which reflects the fact that twists can act as sources and sinks for $\epsilon$ anyons. To study the braiding and fusion of two $\sigma_+$ twists, we define Majorana operators as string operators winding around twists with the same end points, as illustrated  in Fig. \ref{fig-sketch-deform-68}\textbf{C} ({supplementary materials I.E}).  We measure the corresponding string operators and find that $\langle-ic_1c_2\rangle$ can both be negative and positive as shown in Fig. \ref{fig-quantum-gates-68}, which  verifies the nontrivial fusion rule $\sigma_+\times\sigma_+=\mathbf{1}+\epsilon$ for $\sigma_+$ twists. %
		See also {fig. S16}
		for results from processor II, where more fusion rules are demonstrated in a similar fashion. 
		
		{ %
			We mention that, for the experiment carried out on processor I, several boundary qubits are involved in only one or two plaquettes so as to reduce the depth of the quantum circuits for preparing the ground state.  %
			This will increase the degeneracy of the ground states for the stabilizer Hamiltonian, but would not affect our purpose of demonstrating the fusion rules. %
			The reason is that, in our experiment, we use a specific quantum circuit to prepare one of the ground states and a unitary protocol with measurements of only stabilizers involved in the Hamiltonian for the demonstration. As a result, other degenerate ground states is irrelevant and the missing boundary stabilizers will not affect the fusion results. %
			This is confirmed by the agreement between the experimental results and theoretical predictions, and further verified by the {fig. S16},
			where the same fusion results are obtained with a more regular geometry of the chosen 30 qubits and all boundary stabilizers added. 
		}

		\begin{figure}[tbp]
			\includegraphics[width=\linewidth]{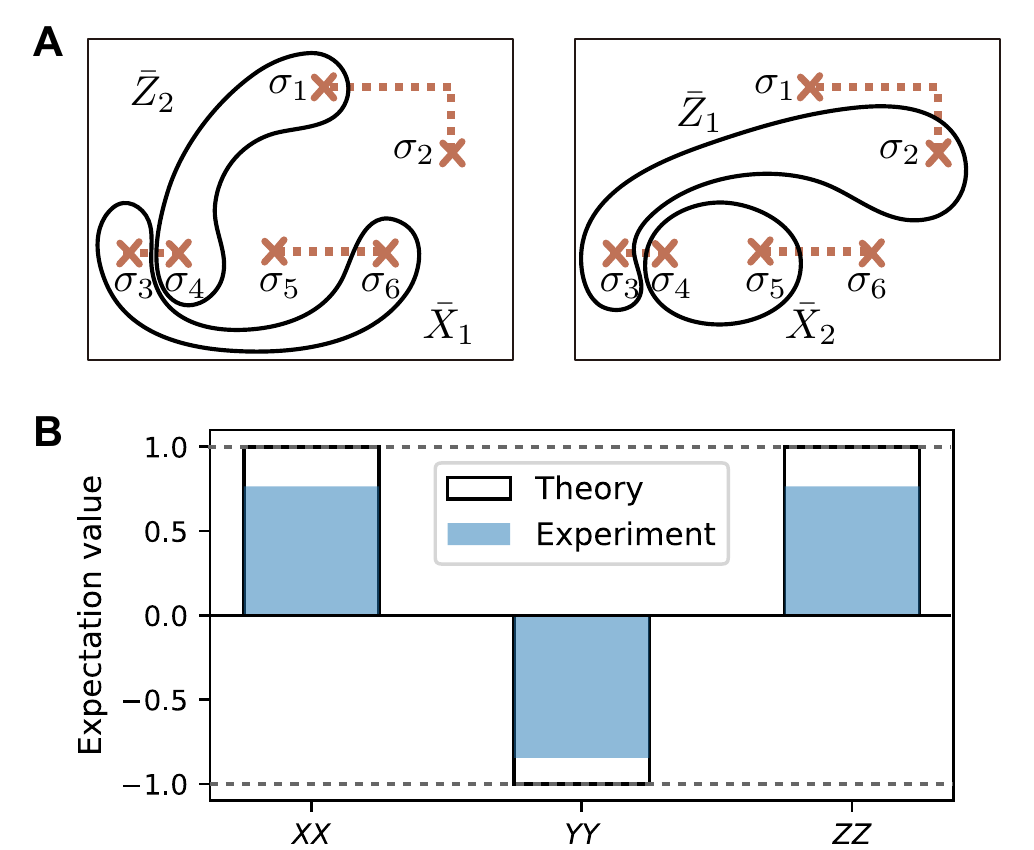}
			\centering
			\caption{\label{fig-bell-68} %
				\textbf{Entangled logical state.} (\textbf{A})  The sketch of the three pairs of twists and the string operators corresponding to the logical ${\bar{Z}_1}$, ${\bar{Z}_2}$, ${\bar{X}_1}$, and ${\bar{X}_2}$ operations (supplementary materials I.I). The encoding scheme is the same as that in Fig. \ref{fig-quantum-gates-68}\textbf{A}. (\textbf{B})  Expectation values of $\bar{X}\bar{X}$, $\bar{Y}\bar{Y}$, and $\bar{Z}\bar{Z}$ for the entangled state of anyon-encoded logical qubits. 
				The overlap between the experimentally prepared state $\rho_\text{exp}$ with the ideal Bell state $|\overline{\Psi}\rangle=(|\overline{00}\rangle+|\overline{11}\rangle)/\sqrt{2}$, defined as $F=\langle\overline{\Psi}|\rho_\text{exp}|\overline{\Psi}\rangle\equiv(1+\langle\bar{X}\bar{X}\rangle-\langle\bar{Y}\bar{Y}\rangle+\langle\bar{Z}\bar{Z}\rangle)/4$, is $0.844\pm 0.002$.
			}\label{Fig-BellState}
		\end{figure}

		\vspace{.5cm}
		\noindent\textbf{\large{}Braidings and logic gates}{\large\par}
		\noindent  Braiding of non-Abelian anyons results in unitary operations in a space spanned by topologically degenerate states. This is the characteristic feature of non-Abelian anyons. As discussed above, we consider six $\sigma_+$ twists  marked in Fig. \ref{GroundStateAndFusion}\textbf{A}. For simplicity and convenience, we omit the symbol ``+" and label them by $\sigma_i$ ($i=1,\cdots,6$), as depicted in  Fig. \ref{fig-quantum-gates-68}\textbf{A}  and Fig. \ref{fig-quantum-gates-68}\textbf{B}.  In general, braidings of twists can be implemented  by  adiabatically transforming the ``geometry" of the Hamiltonian, Majorana tracking \cite{Litinski2017Braiding}, or code deformation \cite{Dennis2002Topological,Raussendorf2007Topological,Bombin2009Quantum}.  Here, we implement braidings of the twists by Majorana tracking  and code deformation (see supplementary materials I.F and I.H).

		In Fig. \ref{fig-quantum-gates-68}, we illustrate how frequently-used single- and two-qubit  gates can be carried out on the encoded logical qubits.   Given that the total topological charge is $\mathbf{1}$, the six $\sigma$ twists can be used to encode two logical qubits. In Fig. \ref{fig-quantum-gates-68}\textbf{A}, we illustrate the encoding scheme used in our experiment. { We first show how single-qubit gates can be implemented for the first logical qubit, through braiding the first three twists.} To this end, in Fig. \ref{fig-quantum-gates-68}\textbf{B} we show the braiding sequence to implement the unitary circuit $\bar{X}\bar{H}\bar{Z}\bar{H}|\bar{0}\rangle$ and our experimental results on measuring the relevant Majorana correlations after the application of each gate. We prepare the system in the logical $|\bar{0}\rangle$ state. We measure the Majorana correlation $\langle-ic_1c_2\rangle$, which corresponds to measuring the fusion charge of $\sigma_1$ and $\sigma_2$. We find $\langle-ic_1c_2\rangle=0.741\pm0.010$, which is consistent with the fact that the system is in the logical $|\bar{0}\rangle$ state initially. A Hadamard gate on the first logical qubit can be applied by sequential braidings of $\sigma_1$ and $\sigma_2$, $\sigma_2$ and $\sigma_3$, and then $\sigma_1$ and $\sigma_2$ again. 
		{This process is in turn equivalent to changing fusion basis, which corresponds to $F$- and $R$-moves in the language of topological quantum computing \cite{nayak2008non}, and can be verified through measuring $-ic_2c_3$.}
		In our experiment, we obtain $\langle-ic_2c_3\rangle=0.005\pm0.002$, which agrees well with the theoretical prediction of $\langle-ic_2c_3\rangle=0$. The logic $\bar{Z}$ gate and $\bar{X}$ gate can be implemented by applying the string operators $-ic_2c_3$ and  $-ic_1c_2$ on the physical qubits, respectively (supplementary materials I.F). {See also figs. S17 and S18}
		for experimental results from processor II, where the corresponding Majorana operators are explicitly shown and the braidings of twists by code deformation are demonstrated as well. %

		The two-qubit logic controlled-X gate ($\overline{\text{CX}}$) can be implemented by the braiding operations shown in Fig. \ref{fig-quantum-gates-68}\textbf{C}, which in turn is equivalent to a change of the fusion basis under the encoding scheme shown in Fig. \ref{fig-quantum-gates-68}\textbf{A}. In order to demonstrate the action of $\overline{\text{CX}}$ gate, we prepare the system in either logical $|\overline{00}\rangle$ or $|\overline{10}\rangle$ state, and then apply sequentially two $\overline{\text{CX}}$ gates. We measure the corresponding Majorana correlations at each step. Our experimental results are shown in Fig. \ref{fig-quantum-gates-68}\textbf{C}, which agree with the corresponding theoretical values. For the state $|\overline{00}\rangle$, applying the $\overline{\text{CX}}$ gate keeps it unchanged and thus the measured Majorana correlations all remain positive. In contrast, for the state $|\overline{10}\rangle$, applying the $\overline{\text{CX}}$ gate evolves it to $|\overline{11}\rangle$, resulting in a sign flip for the relevant Majorana correlations. 
		The deviation of the measured results from the ideal theoretical predictions is due to experimental imperfections and the fact that the measurements of the Majorana correlations are multiqubit nonlocal measurements.

		\vspace{.5cm}
		\noindent\textbf{\large{}Entangled logical states}{\large\par}	
		\noindent  With the logic gates discussed above, we can prepare an entangled state for the logical qubits. To demonstrate this in experiments, we first prepare the system on the logical $|\overline{00}\rangle$ state in the same way as in Fig. \ref{fig-quantum-gates-68}\textbf{C}. We then apply the logic Hadamard and $\overline{\text{CX}}$ gates to evolve the state to the logical Bell state $|\overline{\Psi}\rangle=(|\overline{00}\rangle+|\overline{11}\rangle)/\sqrt{2}$. 
		In Fig. \ref{Fig-BellState}\textbf{A}, we sketch the string operators corresponding to the logical ${\bar{Z}_1}$, ${\bar{Z}_2}$, ${\bar{X}_1}$, and ${\bar{X}_2}$ operations. We measure the expectation values of $\langle\bar{X}\bar{X}\rangle$, $\langle\bar{Y}\bar{Y}\rangle$, and $\langle\bar{Z}\bar{Z}\rangle$, and the results are shown in Fig. \ref{Fig-BellState}\textbf{B}. From this figure, we obtain that the fidelity between the experimentally prepared logical state $\rho_{\text{exp}}$ and the ideal Bell state $|\overline{\Psi}\rangle$ is  $0.844\pm 0.002$, which is larger than $0.5$, indicating that the two logical qubits are entangled.  We mention that, due to the particular encoding scheme used in our experiment, the procedure of preparing the Bell state in our experiment is in fact equivalent to a change of the fusion basis. %
		
		{
			Similarly, we exploit eight twists, four created by code deformation and the other four from  the naturally existed ones at the corners, to encode three logical qubits and prepare a logical GHZ state on quantum processor II ({fig. S19}).
			In {fig. S19}\textbf{A},
			we show the locations of the eight twists used and the corresponding string operators for logical ${\bar{Z}_1}$, ${\bar{Z}_2}$, ${\bar{Z}_3}$, ${\bar{X}_1}$, ${\bar{X}_2}$, and ${\bar{X}_3}$ observables. With these logical operators, we perform quantum state tomography for the prepared logical state and the result is displayed in {fig. S19}\textbf{B},
			from which a fidelity of $0.771 \pm 0.004$ is obtained.

		}

		\vspace{.5cm}
		
		{
			\noindent \textbf{\large{}Discussion and outlook}{\large\par}
			\noindent 	 Although the braiding and fusion properties of twists resemble that of Ising anyons,  they cannot be regarded as non-Abelian Ising anyons \cite{Bombin2010Topological}. They are not intrinsic, finite-energy excitations of the system, and the unitary operation generated by braiding them is topologically protected only up to a nonuniversal overall phase. Thus, strictly speaking, the braiding statistics of the extrinsic twists is only well defined up to a phase, which is referred to as projective non-Abelian statistics in the literature \cite{Barkeshli2013Twist,You2012Projective}. We also clarify that all measurements in our experiment are destructive and the protocols are carried out without quantum error correction and thus are not endowed with topological protection in the strict sense. Achieving this requires consecutive non-destructive stabilizer measurements during the braidings of the twists, which is exceedingly challenging with a system size as large as 68 qubits for the state-of-the-art quantum technologies. %
			
			The digital approach explored in our experiment is highly flexible and generally applicable to simulate a wide range of non-Abelian anyons. Recently, it has been shown theoretically that a broad family of non-Abelian states with a characterizing Lagrangian subgroup can be created efficiently through moderate-depth quantum circuits plus a single measurement layer \cite{Tantivasadakarn2022Shortest}. This is within the reach of the current quantum technologies and it would be interesting to realize such non-Abelian topological orders and study their peculiar properties with programmable quantum processors as shown in this work. In addition, our digital simulation approach carries over readily to realizations of 
			unconventional Floquet topological phases \cite{Zhang2022Digital,Dumitrescu2022Dynamical}, which also paves the way to simulating non-Abelian anyons and studying their unusual features in dynamically driven systems, such as these hosted by the Floquet color code \cite{Kesselring2022Anyon} or Floquet spin liquids \cite{Kalinowski2022NonAbelian}.

		}

		\vspace{.5cm}
		\noindent\textbf{\large{}Data availability}
		{The data presented in the figures and that support the other findings of this study are available for download at
			\href{https://doi.org/10.5281/zenodo.7905826}{https://doi.org/10.5281/zenodo.7905826}}.

		\vspace{.5cm}
		\noindent\textbf{Acknowledgement} We thank L.M. Duan, X. Gao, S.T. Wang, and D. Yuan for helpful discussion.
		The device was fabricated at the Micro-Nano Fabrication Center of Zhejiang University.  We acknowledge the support of the National Natural Science Foundation of China (Grants No. 92065204, 12075128, T2225008, 12174342, 12274368, 12274367, U20A2076, and 11725419), {Innovation Program for Quantum Science and Technology (Grant No. 2021ZD0300200),} and the Zhejiang Province Key Research and Development Program (Grant No. 2020C01019). 
		Z.-Z.S., W.L., W.J., and D.-L.D. are supported by Tsinghua University, and the Shanghai Qi Zhi Institute.

		\vspace{.3cm}
		\noindent\textbf{Author contributions}  S.X., K.W. and L.X. ~carried out the experiments under the supervision of C.S., Q.G.~and H.W.. J.C. and X.Z. designed the device and H.L. fabricated the device, supervised by H.W.. Z.-Z.S.~designed the quantum circuits under the supervision of D.-L.D.. W.L., W.J., L.Y., Z.-Z.S., and D.-L.D.~conducted the theoretical analysis. All authors contributed to the experimental set-up, analysis of data, discussions of the results, and writing of the manuscript.

		\vspace{.3cm}
		\noindent\textbf{Competing interests}  All authors declare no competing interests.
		\let\oldaddcontentsline\addcontentsline
		\renewcommand{\addcontentsline}[3]{}

\let\addcontentsline\oldaddcontentsline
	
\clearpage
\newpage 
\onecolumngrid
\setcounter{section}{0}
\setcounter{equation}{0}
\setcounter{figure}{0}
\setcounter{table}{0}
\makeatletter
\renewcommand\thefigure{S\arabic{figure}}
\renewcommand\thetable{S\arabic{table}}
\renewcommand\theequation{S\arabic{equation}}

\begin{center} 
	{\large \bf Supplementary Materials: Digital simulation of non-Abelian anyons with 68 programmable superconducting qubits}
\end{center} 

\setcounter{figure}{0}
\setcounter{table}{0}
\renewcommand\thefigure{S\arabic{figure}}
\renewcommand\thetable{S\arabic{table}}
\maketitle
\tableofcontents

\section{Theory}
\label{app:theory1}

\subsection{Non-Abelian anyons}

It is well recognized that bosonic and fermionic statistics are the only two allowed quantum statistics in three-dimensional (3D) space where we live in. There is only one topologically distinct way to exchange two identical particles in 3D space. In such a scenario, the many-body wave functions of identical particles host only two possible symmetries under exchange: the symmetric wave functions for bosons, and the anti-symmetric wave functions for fermions. However, such a  boson-fermion dichotomy can be broken in 2D space, where the swap of two identical particles may give rise to a topological non-trivial path, and thus the exotic anyon statistics would emerge.

From the perspective of group representation theory in mathematics, the bosonic and fermionic statistics are described by the two one-dimensional irreducible representations of the permutation group, whereas the anyonic statistics are described by the braid group. According to the representation of braid group, anyons can be divided into two categories, Abelian and non-Abelian anyons. Abelian anyons usually correspond to the one-dimensional irreducible representations of braid group, {\it i.e.}, all of the braid group elements commute.  Whereas non-Abelian anyons usually correspond to the higher dimensional irreducible representations of the braid group, and the braid elements multiplication is non-commutative.  

In physics, the study of  Abelian anyons dates back more than forty years ago. For demonstration,  let us suppose that we have $N$ identical particles with the wave function $|\Psi(r_1,r_2,\cdots r_N)\rangle$, where $r_i$'s denote the positions of identical particles in real space.  Exchanging the position of arbitrary two particles could give rise to an overall phase 
\begin{equation}\label{anyon_statistics}
	|\Psi(r_1,r_2,\cdots r_N)\rangle\rightarrow e^{i\theta}|\Psi(r_1,r_2,\cdots r_N)\rangle.
\end{equation} 
The special cases $\theta = 0$ or $\pi$ correspond to the bosonic or fermionic statistics. While for other values of the statistical angle $\theta \in [0,2\pi)$ except $0$ or $\pi$, the corresponding identical particles are called Abelian anyons.

The Abelian anyons are expected to appear in a variety of two-dimensional quantum systems. For example, in the $\nu=1/3$ fractional quantum Hall system, there appear localized excitations (e.g. quasiparticles or quasiholes) carrying quantized magnetic fluxes and fractional charges. Such localized excitations obey Abelian anyonic statistics, since exchanging the corresponding two composite particles would give rise to a non-trivial phase owing to their mutual Aharonov-Bohm effect. Thus, the localized excitations can be regarded as Abelian anyons.  Besides that, the Abelian anyonic excitations are also predicted in the lattice model, like the Kitaev $\mathbb{Z}_2$ toric code model with ground state degeneracy depending on the topological structure of the two-dimensional system. In such $\mathbb{Z}_2$ toric model, the pairs of Abelian anyons $e(m)$ can be excited by applying the Pauli-$Z(X)$ operator on the local lattice site. By applying the string operator on the excited states, we can move the corresponding anyons on the lattice. Encircling the $e$-anyon around the $m$-anyon results in a $e^{i\pi}$ phase factor to the corresponding state, rather than the phase factor $1$ for bosons or fermions. This behaves the Abelian anyonic statistics. However, braiding Abelian anyons in the subspace spanned by the degenerate ground states cannot be utilized for quantum computation, since the braiding operation would only induce an overall phase factor to the corresponding state. %

Different from the exchange of Abelian anyons, exchanging non-Abelian anyons can give rise to a unitary operation on the degenerate ground state space. For example, suppose we have a set of degenerate ground states $\{|\Phi_i(r_1,r_2,\cdots r_N)\rangle\}$ of a topological quantum system with $N$ identical non-Abelian anyonic excitations. Then by exchanging the positions of two non-Abelian anyons, one obtains
\begin{equation}
	|\Phi_i(r_1,r_2,\cdots r_N)\rangle\rightarrow U_{ij}|\Phi_j(r_1,r_2,\cdots r_N)\rangle,
\end{equation}
where $U$ denotes the unitary operation. The study of non-Abelian anyons has attracted extensive interest in both theory and experiments, owing to not only their exotic physics but also their potential applications in topological quantum computation against local noises. The non-Abelian anyons are predicted to appear in $\nu=\frac{5}{2}$ fractional quantum Hall system, quantum double models of non-Abelian finite group, and so on. One of the simplest non-Abelian anyon models is the Ising anyon model,  which is described by the $SU(2)_2$ topological quantum field theory.  The Ising anyon model includes three anyons, the vacuum $\bf{1}$, the Ising anyon $\sigma$, and the Majorana fermion $\epsilon$. They obey the following fusion rules
\begin{equation}
	\begin{aligned}
		\label{eq-fusion-Ising}
		&\sigma\times\bf{1}=\bf{1}\times\sigma= \sigma, \\
		&\epsilon\times\bf{1}=\bf{1}\times\epsilon= \epsilon,\\
		&\sigma\times\epsilon=\epsilon\times\sigma= \sigma,\\
		&\epsilon\times\epsilon=\bf{1},\\
		&\sigma\times\sigma=\bf{1}+\epsilon. 
	\end{aligned}
\end{equation}
For a system with  $2N$ Ising anyons fusing to the vacuum, the total Hilbert space spanned by the fusion bases is $2^{N-1}$. Thus the quantum dimension of the Ising anyon is $d=\sqrt{2}$. Applying the braiding operations on the Ising anyons, one can thus obtain the $2^{N-1}$ -- dimensional irreducible representation of the $2N$-strand braid group $\mathcal{B}_{2N}$. From the perspective of quantum computation, the fusion bases of $2N$ Ising anyons span the full computational space, and the elements of the braid group serve as the quantum logic gates. Thus, the universality of the topological quantum computation model based on braiding operations would depend on the universality of the braid group in expressing the $SU(2^{N-1})$ group.

One candidate to realize the Ising anyon statistics is based on the Majorana zero modes (MZM). In recent years, extensive experimental efforts have been devoted to demonstrating the presence of the MZM in quantum materials. However, it remains a  challenging task to confirm the existence of MZM in a solid-state system, as well as braiding such MZMs. 

As mentioned above, the non-Abelian anyonic excitations can also emerge in the quantum double model of non-Abelian finite group, where the dimension of each local lattice site is equivalent to the group size.  The simplest quantum double lattice model of the non-Abelian finite group is $D(S_3)$, where $S_3$ denotes the 3-strand symmetric group. The eight anyonic excitations correspond to the eight irreducible representations of the Hopf algebra $\mathcal{D}(S_3)$. In a recent work \cite{Bravyi2022Adaptive}, the authors show that the ground states and anyon excitations of the model $D(S_3)$ can be realized by adaptive circuits with constant-depth,  assisted by those geometrically local unitary gates and mid-circuit measurements.

\subsection{Characterization of the model Hamiltonian}

As we have introduced in the main text, the Hamiltonian of the toric code is a sum of stabilizers. Any two of these stabilizers share an even number of common qubits attached to different Pauli operators. Thus, all the stabilizers commute with each other and can be simultaneously diagonalized. They restrain the ground state of the toric code to satisfy ${{\bf{A}}_{\bf k}}\left| G \right\rangle  = \left| G \right\rangle $ for all $\mathbf{k}$. Each restriction $\left\langle {{{\bf{A}}_{\bf k}}} \right\rangle  = 1$ will halve the number of free parameters to describe the ground state. For a system with $N$ qubits, the dimension of the corresponding Hilbert space is ${2^N}$. Its ground state can be uniquely determined by a complete set with $N$ independent stabilizers, up to an irrelevant global phase \cite{Fowler201209Surface}.

The deformation of the lattice may reduce the number of independent stabilizers and thus bring extra degeneracy to the ground states. For example, the deformation area of the lower left corner in Fig. 2(a) of the main text is the product of two stabilizers ${{\bf{A}}_{\bf k}}$, written as ${{\bf{A}}_{{D_3}}} = {{\bf{A}}_{\left( {9,1} \right)}}{{\bf{A}}_{\left( {9,2} \right)}}$. As a result, the adding of deformation $D_3$ will reduce the number of independent stabilizers by one.  Another way to define the stabilizer for a deformation is to attach Pauli-$Y$ operators to qubits on its edge,  as shown in Fig. \ref{fig-ground-state-30}. The lattice sites shared by three (rather than four) neighboring plaquettes are defined as twists, which must appear in pairs. The existence of one pair of twists reduces the number of independent stabilizers by one, resulting in double degeneracy of the ground states. This double degenerate ground states can be used to define the two possible fusion results of Ising anyons. Generally speaking, one pair of twists correspond to the double degeneracy of the ground states, and can be used to mimic the behavior of a pair of Ising anyons whose quantum dimension is two. On quantum processor I, the imperfect geometry for the chosen $68$ qubits naturally introduces twists that mimic Ising anyons under braiding and fusion. The braiding of non-Abelian {twists} and the logic quantum gates are performed on the ground-state space where the stabilizers in Fig. 2(a) are all positive. We mention that these are not a complete stabilizer set for 68 qubits and not all the degenerate ground states are employed.

\subsection{String operators}

In this work, observables and unitary operations are all described by string operators. For concreteness, we take the quantum processor II as an example to explain the quantum circuit implementation corresponding to the string operators. The chosen {$30$} qubits on quantum processor II form a regular rectangular lattice (Fig. \ref{fig-toric-code-30}), so all string operators can be displayed more succinctly and conveniently. The explicit definition rules for stabilizers on the deformed lattice are exemplified in Fig. \ref{fig-ground-state-30}. In addition, we also use the Zig-Zag index for simplicity and convenience: for ${\bf k} = \left( {a,b} \right)$, we use $ k^{\prime} = (a-1) \times col + b$ as the index, where $col$ denotes the number of columns. 

Now we introduce the constructions and properties of string operators. A string operator is an ordered sequence of Pauli operators defined on the vertices of the lattice, which is a Hermitian Pauli string. The corresponding Pauli operator is ${X}$ if the string operator  is oblique to the right, and ${Z}$ if it is oblique to the left. The string operators may have crosses, which should be attached with both Pauli ${X}$ and ${Z}$. In this work we attach Pauli ${Y}$ to the crosses of string operators to ensure that string operators are Hermitian.

One segment of the string operator can flip the sign of two adjacent stabilizers on its diagonal. The Pauli ${X}$ on position ${\bf k}$ flips the sign of ${{\bf{A}}_{\bf{{k} - {i}}}}$ and ${{\bf{A}}_{\bf{{k} - {j}}}}$. It can create or annihilate excitations in pairs. Meanwhile, the Pauli ${Z}$ on position ${\bf k}$ flips the sign of ${\bf{A}_{{k} - {i} - {j}}}$ and ${{\bf{A}}_{\bf k}}$. The string operator flips the sign of stabilizers on its starting and ending positions, which is illustrated  in Figs. \ref{fig-toric-code-30}\textbf{B} and \textbf{C}. Particularly, a closed string operator does not influence the sign of all ${{\bf{A}}_{\bf k}}$. It commutes with the Hamiltonian and can be implemented on the ground state without resulting in any excitations. The closed string operators have several important properties. A closed string operator is a Hermitian Pauli string commuting with all ${{\bf{A}}_{\bf k}}$. It is taken as a stabilizer to be measured and characterizes different degenerate ground states. The string operator is also unitary, so it can be represented by a quantum circuit and applied to a physical state. As a result, string operators are used to describe both logical Pauli operators and logical observables in this work.

In a square lattice with no deformation, string operators create or move the charge $e$ and $m$ but cannot change their charge types. Closed string operators  act trivially  and generate a charge of vacuum. To demonstrate the non-Abelian statistics, we change the ``geometry'' of Hamiltonian on quantum processor II. The new stabilizers and experimental results of the ground state are shown in Fig. \ref{fig-ground-state-30}. The deformation introduces four twists that exhibit non-Abelian statistics.  When deformations exist, the string operator can start at $e$ and end at $m$ if it winds around an odd number of twists. This means that the charge $e$ is transformed to charge $m$. Whereas, the charge $e$ is preserved after surrounding an even number of twists. These statements also apply to charge $m$. The effect of string operators that cross different number of twists is illustrated in Fig. \ref{fig-exchange-charge-30}.

\subsection{Charge of Ising-type twists}

The charges of Abelian anyons are identified by the emergent global phase when braiding them. For example, whether there are an odd number of $e$ anyons can be detected by circling this area and returning to its original position with an $m$ anyon. The global phase of -1 indicates the existence of an odd number of $e$ anyons. An interferometric measurement can identify the phase, but it might be challenging experimentally. On the other hand, the string operators that drag anyons around are Hermitian for $e$, $m$, and $\epsilon $. Thus, we can directly measure the string operator to distinguish these charges from the vacuum. 

The case of charge $\sigma $ is a bit complicated since the twist can change the charge \cite{Bombin2010Topological}. An $e$ anyon cannot return to its original position after winding around a twist. One solution is to circle the twist with another twist. {This requires geometric deformation of the lattice thus challenging in experiments.} Another solution is to circle the twist twice to close the corresponding string operator. This scheme can be realized with a bunch of single-qubit gates since the string operator is composed of Pauli operators, as shown in Figs. 1\textbf{B} and \ref{fig-fusion-30}\textbf{A}. Note that string operators circling an odd number of twists have eigenvalues of $ \pm i$. The complex number emerges at the cross of ${{S}_\sigma }$, where Pauli ${Z}$ and ${X}$ act on the same qubit successively, noting ${ZX} = i{Y}$. To ensure that the string operators are Hermitian, we attach Pauli ${Y}$ to the crosses of string operators instead of $ZX$. Thus, the string operators to be measured are Hermitian Pauli sequences of $ - i{S_\sigma }$.

Like other string operators characterizing topological charges, {those strings corresponding to the charge of twists can be transformed to topologically equivalent ones.} Specifically, they can be arbitrarily multiplied by the stabilizers in the Hamiltonian. The resulting string operators characterize the charge of twists in different areas. {In Fig. \ref{fig-fusion-30}\textbf{A}, we show the shortest string operators that can characterize the charge of four twists on the quantum processor II.} The Pauli strings of these shortest Ising charge operators are the same as the stabilizers on the corresponding areas, which is consistent with the cases of charges $e$, $m$, and $\epsilon$. Some fusion rules from Eq. \ref{eq-fusion-Ising} are verified on the quantum processor II. The corresponding experimental results are shown in Fig. \ref{fig-fusion-30}\textbf{B}.

\subsection{Majorana operators}

Anyons with non-Abelian statistics can be used to perform topological quantum computing \cite{nayak2008non}. One way to determine whether given anyons bear non-Abelian statistics or not is to study the dimension of their fusion space. The anyons with definite fusion result are Abelian anyons. For example, the charges $e$, $m$ and $\epsilon $ considered in this work only have trivial fusion space, thus are Abelian. The braiding sequence of these anyons can be exchanged arbitrarily and cannot be used for quantum computing. However, if we consider a subset of the charges $\left\{ {{\bf{1}},{\sigma _ + },\epsilon } \right\}$, their fusion rules from Eq. 2 are exactly the fusion rules of Ising anyons. One promising topological quantum computing scheme is to implement logic quantum gates by braiding the Ising anyons, and to access the measurements of logical qubits with the fusion results \cite{nayak2008non}.

In this work, the fusion and braiding of Ising-type twists are described by Majorana operators, which are in turn defined as string operators across different twists. Four Majorana operators on quantum processor II are shown in Fig. \ref{fig-logical-gates-30}\textbf{A}. All these Majorana operators start and end at the same positions, which is used to ensure that the product operator of any two of them is a closed string operator that commutes with the system Hamiltonian. Since two different Majorana operators start and end at the same positions, they have to cross an odd number of times. Given the definition of string operators, two Majorana operators always anti-commute with each other. Taking Majorana operators ${c_1}$ and ${c_2}$ shown in Fig. \ref{fig-logical-gates-30}\textbf{A} as an example, we have $c_1={Z_{14}}{X_{19}}{Z_{25}}{X_{26}}$ and $c_2={Z_{20}}{X_{14}}{Z_8}{Z_7}{X_1}{Z_2}{Z_9}{X_{15}}$. One can directly verify that they are anti-commutative ${c_1}{c_2}=- {c_2}{c_1}$. The anti-commutativity of ${c_1}$ and ${c_2}$  comes from the anti-commutativity of ${{Z_{14}}}$ in $c_1$ and ${{X_{14}}}$ in $c_2$.

The fusion result of two {twists} can be uniquely identified by measuring the corresponding Majorana correlator. It can either be ${\bf{1}}$ or $\epsilon $ according to Eq. 2. Since both $e$ and $m$ anyons generate a phase factor of $\pi $ (0) after winding around an $\epsilon $ charge (the vacuum $\mathbf{1}$), the product of two Majorana operators that drag $e$ or $m$ anyon around two twists can characterize the fusion results of two {twists} in the subset of charges $\left\{ {1,{\sigma _ + },\epsilon } \right\}$ \cite{Bombin2010Topological}.

The braiding of Ising anyons can be regarded as the braiding of Majorana operators, which changes their indices and phase factors \cite{nayak2008non}. For example, {the braiding of ${\sigma _1}$ and ${\sigma _2}$ (denoted as ${B_{12}}$) can be written as \cite{Ogburn1999Topological}}
\begin{align}\label{eq-Majorana-braiding}
	{c_1}\buildrel {{B_{12}}} \over
	\longmapsto {c_1}{c_2}c_1^{ - 1},\; {c_2}\buildrel {{B_{12}}} \over
	\longmapsto {c_1}.
\end{align}
Noting that the Majorana operators are fermionic $\left\{ {{c_1},{c_2}} \right\} = 0$, one can obtain ${c_1}{c_2}c_1^{ - 1} =  - {c_2}$.

\subsection{Logical qubits, gates, and measurements}
\label{Logical qubits, gates, and measurements}
\begin{figure}[htb]
	\includegraphics[width=1\linewidth]{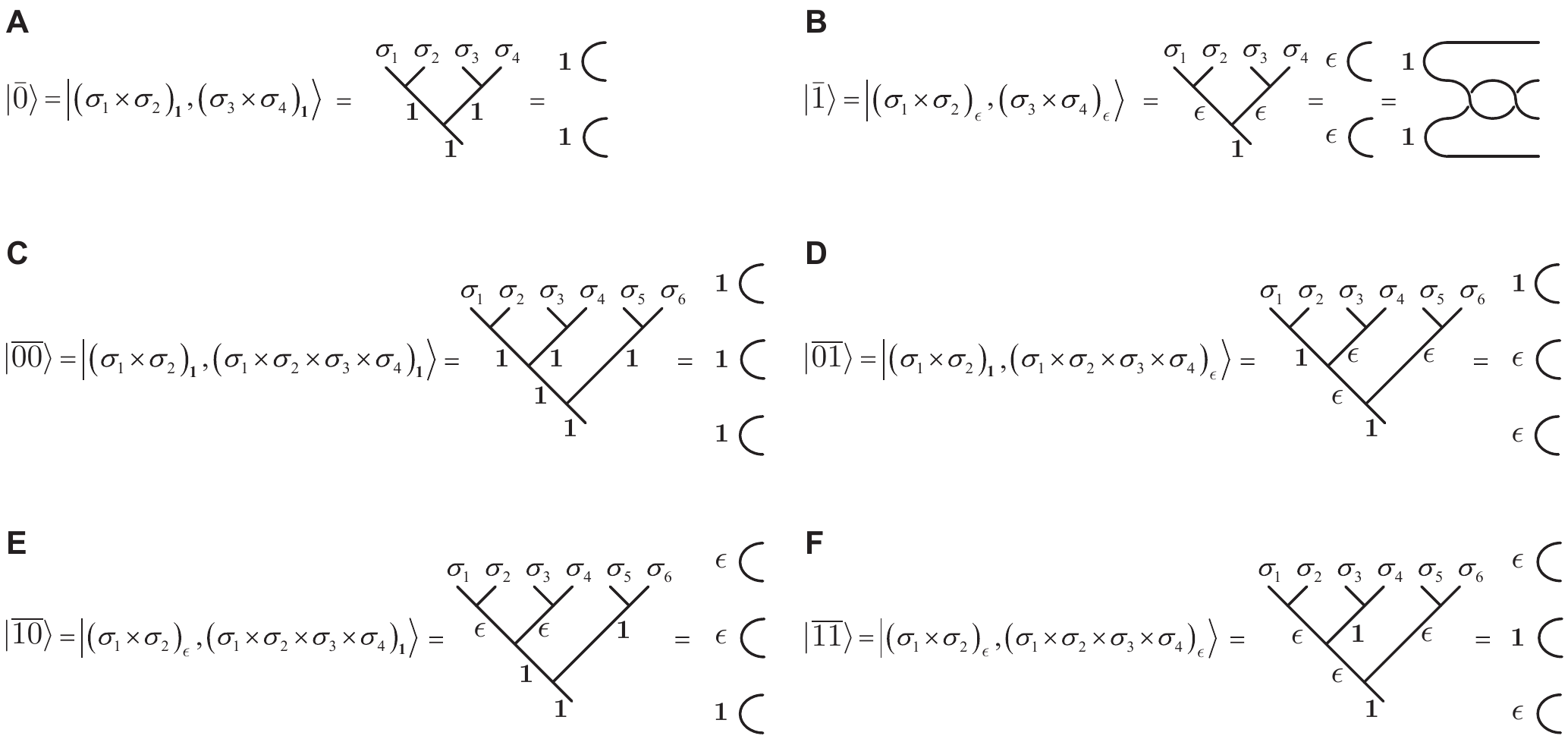}
	\caption{\label{fig-logical-qubits} Dirac notations, fusion trees, and braiding sequences of logical qubit states. (\textbf{A})  $| {\bar 0} \rangle $ represents logical $\left| {0} \right\rangle $ state. { This state is characterized by that the fusion outcome of the first two twists is the vacuum, as shown in the Dirac notation and fusion tree notation.} The braiding sequence on the right shows that these two pairs of {twists} are both generated from the vacuum, which is consistent with their fusion results. (\textbf{B})  Logical state $| {\bar 1} \rangle $. The braiding sequence of $| {\bar 1} \rangle  = {\bar X}| {\bar 0} \rangle $ is also shown. ({\textbf{C})  Logical state $| \overline{00} \rangle $. The first logical qubit being logical state $| {\bar 0} \rangle $ is characterized by that the first two twists fuse to the vacuum. And the second logical qubit being $| {\bar 0} \rangle $ is characterized by that the first four twists fuse to the vacuum. The second logical qubit can also be characterized by the fusion results of the last two twists, due to the conservation of total topological charge.} (\textbf{D})  Logical state $| \overline{01} \rangle $. (\textbf{E})  Logical state $| \overline{10} \rangle $. (\textbf{F})  Logical state $| \overline{11} \rangle $.}
\end{figure}

{We can define the logical qubits in the fusion space of Ising anyons. The number of Ising anyons required can be obtained by the degrees of freedom related to fusion outcomes. Considering that the quantum dimension of a pair of Ising anyons is two, the quantum dimension of two pairs of Ising anyons is enough to define a logical qubit and three pairs of Ising anyons can define two logical qubits. In this work, we define the case of the first two twists (corresponding to two Ising anyons) fusing to the vacuum as the logical $\left| 0 \right\rangle $ state, denoted as $\left| {\bar 0} \right\rangle $. The case of the first four twists fusing to the vacuum is defined as $\left| {\bar 0} \right\rangle $ state of the second logical qubit.} There are several notations for the logical states in the main text to facilitate writing and vividly illustrate the braidings. We show a complete set of the logical bases represented by these notations in Fig. \ref{fig-logical-qubits}.

\begin{figure}[htb]
	\includegraphics[width=1\linewidth]{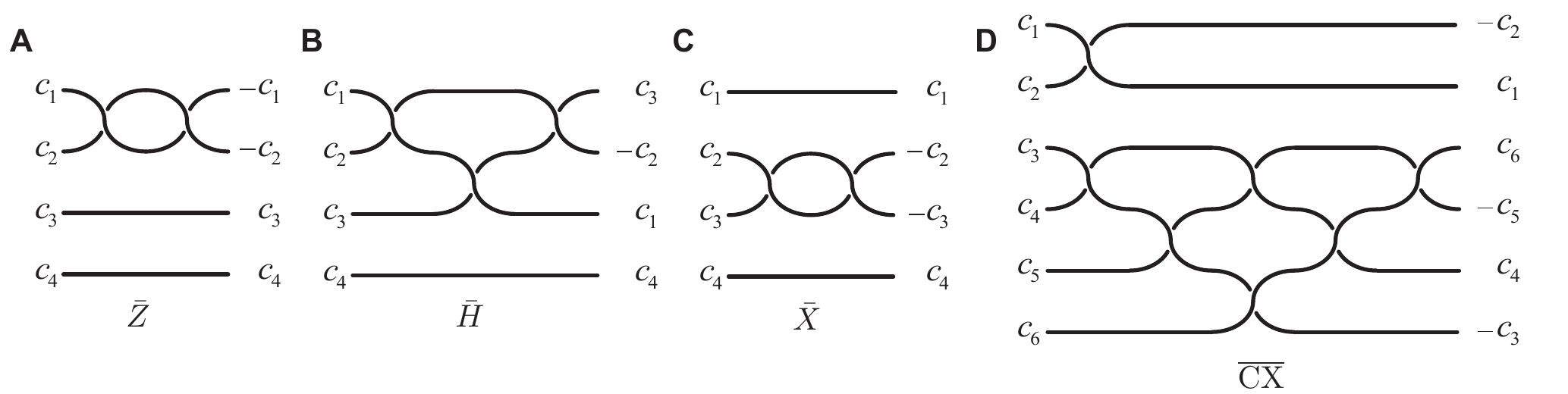}
	\caption{\label{fig-braid-Majorana} The braiding of Majorana operators corresponding to all logical operators in this work. (\textbf{A})  $\bar Z = B_{12}^2$. Braiding the first two {twists} twice gives ${c_1} \mapsto  - {c_1}$, ${c_2} \mapsto  - {c_2}$, ${c_3} \mapsto {c_3}$, and ${c_4} \mapsto {c_4}$, corresponding to logical $\bar Z$ gate. (\textbf{B})  $\bar H = {B_{12}}{B_{23}}{B_{12}}$. (\textbf{C})  $\bar X = B_{23}^2$. (\textbf{D})  $\overline {\text{CX}}  = {B_{12}}{B_{34}}{B_{45}}{B_{34}}B_{56}^{ - 1}B_{45}^{ - 1}B_{34}^{ - 1}$.}
\end{figure}

For a given encoding scheme and quantum gate, calculating the corresponding braiding sequence is complicated but eminently solvable \cite{Nayak200809RoMP, Georgiev200802Towards}. In Fig. \ref{fig-braid-Majorana}, we show all the explicit braiding sequences and their effects on Majorana operators for the logic gates ${\bar Z}$, ${\bar X}$, ${\bar H}$, and $\overline {{\text{CX}}} $.
The scheme of exploring Majorana operators and changing fusion basis is used to implement the logic $\bar{H}$ and $\overline{\text{CX}}$ gates in our experiments. As pointed out in Ref. \cite{Litinski2017Braiding}, Majorana tracking is braiding and can be used to probe the non-Abelian statistics.

Braiding Majorana operators can in turn be implemented by unitary gates acting on the physical qubits \cite{Bravyi2006Universal}. For example, the logical $\bar{X}$ operator can be realized by the braiding sequence of $B_{23}^2$, where the logical qubit is defined by four {twists}, as shown in Fig. 3(b). The measurement of the logical qubit under logical $\bar{Z} $ basis is the fusion result of the first two twists, which can be obtained by measuring $ - i{c_1}{c_2}$. After the logic $\bar{X}$ gate, the Majorana operator changes as ${c_2} \mapsto  - {c_2}$ and ${c_3} \mapsto  - {c_3}$. The unitary gate corresponding to this process is ${{{U}}_{\bar X}} =  - i{c_2}{c_3}$. One can verify that ${{U}}_{\bar X}^{ - 1}{c_2}{{{U}}_{\bar X}} =  - {c_2}$ and ${{U}}_{\bar X}^{ - 1}{c_3}{{{U}}_{\bar X}} =  - {c_3}$, which realizes the logic $\bar{X} $ gate. Similarly, the logic $\bar{Z} $ gate is defined as ${{{U}}_{\bar Z}} =  - i{c_1}{c_2}$.

Here for clarity we give a more specific description of how to implement logic quantum gates in the context of Fig. 3(b). In this figure, four Majorana operators $c_1$, $c_2$, $c_3$, and $c_4$ are used to define the logic gates and observables (Majorana correlations). Initially, the string operator corresponding to the logic $\bar{Z}$ gate is ${{{U}}_{\bar Z}} =  - i{c_1}{c_2}$. The first logic $\bar H $ gate is implemented by changing the Majorana operators as ${c_1} \mapsto {c_3}$, ${c_2} \mapsto {-c_2}$, ${c_3} \mapsto {c_1}$, and ${c_4} \mapsto {c_4}$. So the string operators defined by Majorana operators are changed accordingly. For example, the unitary to implement the logic $\bar{Z}$ gate changes to $ - i{c_2}{c_3}$, which is obtained by transforming ${{{U}}_{\bar Z}}$ according to ${c_1} \mapsto {c_3}$ and ${c_2} \mapsto {-c_2}$. The Majorana correlator to be measured as logical $\bar{Z}$ observable also changes to $ - i{c_2}{c_3}$ at the same time.
The logic $\bar{Z}$ gate is implemented by applying the unitary gate $ - i{c_2}{c_3}$ on physical qubits, which changes neither the indices nor phase factors of the Majorana operators. 
The second logic $\bar{H}$ gate changes the Majorana operators as ${c_3} \mapsto {c_1}$, ${-c_2} \mapsto {c_2}$, ${c_2} \mapsto {c_3}$, and ${c_4} \mapsto {c_4}$, where all Majorana operators recover their original indices and phase factors. 
The last logic $\bar{X}$ gate is implemented by ${{{U}}_{\bar X}} =  - i{c_2}{c_3}$, considering the recovery of Majorana operators.

In the above discussion, we have defined the logical qubits and logic gates in the fusion space of twists. The measurement of logical qubits reduces to measuring the fusion results. Thus, the preparation and characterization of the logical Bell state is accessible in the experiment. We define two logical qubits in the fusion space of three pairs of twists. The corresponding charges of fusing three pairs of twists are all initialized to vacuum, where the initial logical state is $\left| {\overline{00}} \right\rangle$. Then we implement the braiding sequence on Majorana operators corresponding to a logic Hadamard gate on the first logical qubit and a logic ${\overline{\text{CX}}}$ gate. The Majorana operators evolve under this braiding sequence according to Eq. \ref{eq-Majorana-braiding}. After the braiding, measuring logical Pauli operators corresponds to measuring the evolved Majorana correlators, which are nonlocal observables. In this way, we can obtain the fidelity of the experimentally prepared logical state.

\subsection{Quantum circuit for the initialization}\label{sec_circuit_for_init}

Here we introduce the method to prepare the ground state of Hamiltonian ${H} =  - \sum\limits_{\bf k} {{\bf{A_k}}} $, where ${\bf{A_k}} = {{X}_{\bf k}}{{Z}_{\bf k + i}}{{Z}_{\bf k + j}}{{X}_{\bf k + i + j}}$, as well as the initialization of the logical states in the fusion space. Initially, the system is in the physical state ${\left| 0 \right\rangle ^{ \otimes N}}$, where $\left\langle {{\bf{A_k}}} \right\rangle = 0 $ for all ${\bf k}$. We aim to prepare the ground states of this Hamiltonian, where all of the local operators ${{\bf{A_k}}}$ share the same eigenvalue of 1. An intuitive initialization method is to perform non-destructive measurements on all ${\bf{A_k}}$ and annihilate the excitations with string operators. This scheme can {prepare} the ground state with only one layer of measurements \cite{Tantivasadakarn202209}. However, performing non-destructive measurements and feed-forward string operators is expensive under the existing experimental conditions.

The non-destructive measurements of ${\bf{A_k}}$ can be achieved by the Hadamard test that requires an ancillary qubit. The procedure is to apply the unitary ${\bf{A_k}}$ on quantum state $\left| \psi  \right\rangle $ of the lattice system conditioned on the ancillary state of ${1 \over {\sqrt 2 }}\left( {\left| 0 \right\rangle  + \left| 1 \right\rangle } \right)$. Then by  applying a Hadamard gate on the ancillary qubit, we obtain
\begin{align}\label{eq-Hadamard-test-result}
	{1 \over 2}\left[ {\left| 0 \right\rangle \left( {{\bf{I}} + {\bf{A_k}}} \right)\left| \psi  \right\rangle  + \left| 1 \right\rangle \left( {{\bf{I}} - {\bf{A_k}}} \right)\left| \psi  \right\rangle } \right].
\end{align}
Simple calculations show that the measurement result on the ancillary qubit equals the expected value $\left\langle \psi  \right|{{\bf{A}}_k}\left| \psi  \right\rangle $. The projection of $\left| \psi  \right\rangle $ to the state with ${\bf{A_k}} = 1$ can be realized by projecting the ancillary qubit to ${\left| 0 \right\rangle }$. According to Eq. \ref{eq-Hadamard-test-result}, projecting the ancillary qubit to ${\left| 0 \right\rangle }$ is to project the quantum state $\left| \psi  \right\rangle $ to ${{1 \over {\sqrt 2 }}\left( {{\bf{I}} + {\bf{A_k}}} \right)\left| \psi  \right\rangle }$, whose eigenvalue corresponding to ${\bf{A_k}}$ is $+1$.

\begin{figure}[htb]
	\includegraphics[width=1\linewidth]{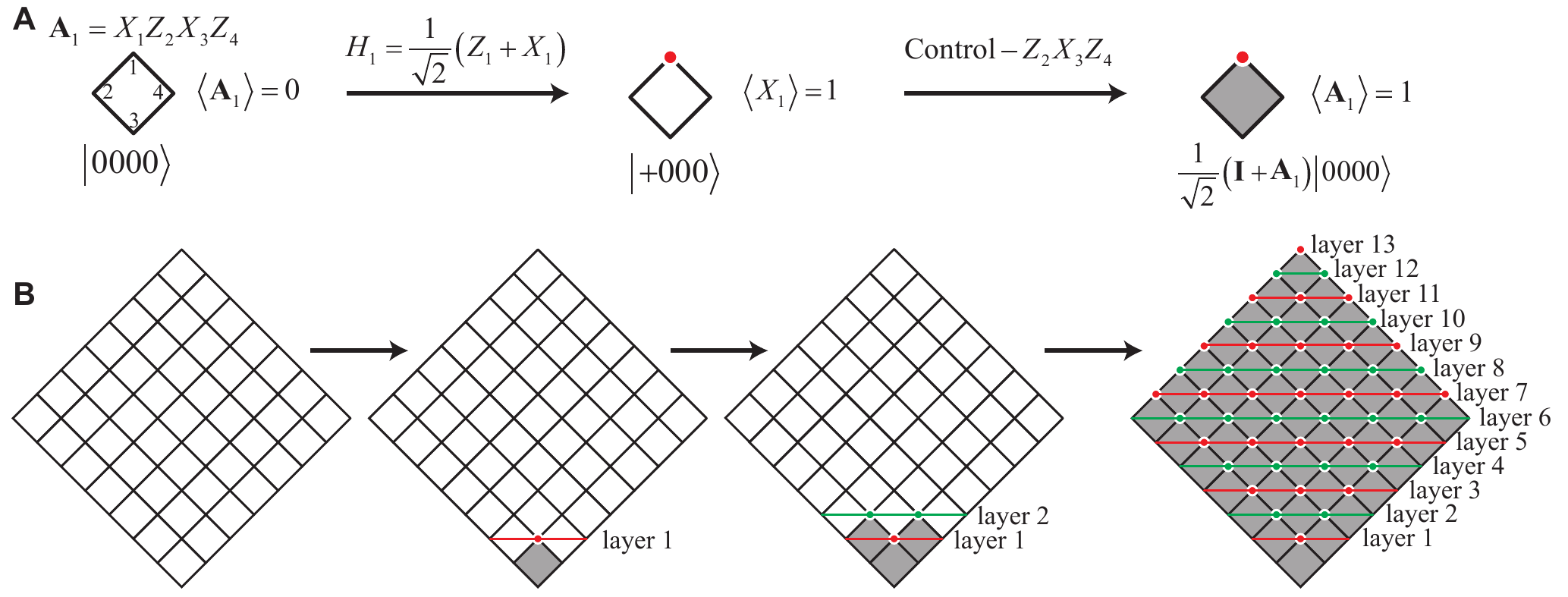}
	\caption{\label{fig-representative-qubits} The method to prepare the ground state of the toric code. (\textbf{A})  The method to prepare the state corresponding to $\left\langle {{{\bf{A}}_1}} \right\rangle  = 1$, where the stabilizer ${\bf{A}}_1 = {X_1}{Z_2}{X_3}{Z_4}$ is defined on one plaquette. The first step is to initialize the representative qubit to a superposition state. The second step is to implement the rest part of ${\bf{A}}_1$ on qubit $2$, $3$, and $4$ controlled by the representative qubit. These controlled operations are $\text{CZ}_2$, $\text{CX}_3$, and $\text{CZ}_4$ that can be sequentially implemented. The final state is the eigenstate of ${\bf{A}}_1$ with the corresponding eigenvalue equal $+1$, which is marked {in} gray. (\textbf{B})  The method to prepare the ground state of the toric code. Starting with the state of ${\left| 0 \right\rangle ^{ \otimes N}}$, we first initialize the plaquette at the bottom, then the second layer above it, and so on. The representative qubits can function well without interfering with each other under the order shown in this figure, where circuit depth is linearly related to the diameter. The colors of representative qubits are only for clarity of layering.}
\end{figure}

The term ${{1 \over {\sqrt 2 }}\left( {{\bf{I}} + {\bf{A_k}}} \right)}$ is not unitary and cannot be directly implemented by quantum circuit. However, we can initialize a representative qubit for each ${{\bf{A_k}}}$ to create the superposition \cite{Liu202211PQ}. Let us use the example of ${{\bf{A}}_1} = {X_1}{Z_2}{X_3}{Z_4}$ with the representative qubit of ${Q_1}$  to show how to implement ${{1 \over {\sqrt 2 }}\left( {{\bf{I}} + {\bf{A_1}}} \right)}$ in detail. Starting from $\left| {0000} \right\rangle $, we firstly initialize the representative qubit to $\left|  +  \right\rangle  = {1 \over {\sqrt 2 }}\left( {\left| 0 \right\rangle  + \left| 1 \right\rangle } \right)$, which is stabilized by ${{X}_1}$. Then we apply the operators of ${{\bf{A}}_1}$ except ${{X}_1}$ controlled by the representative qubit. The resulting state is 
\begin{align}\label{eq-initialize-A1}
	\left| 0 \right\rangle \left| {000} \right\rangle  + {{X}_1}{{\bf{A}}_1}\left| 1 \right\rangle \left| {000} \right\rangle  = {1 \over {\sqrt 2 }}\left( {{\bf{I}} + {{\bf{A}}_1}} \right)\left| {0000} \right\rangle.
\end{align}
Here ${{X}_1}{{\bf{A}}_1}$ is the product of all single-qubit operators in {${{\bf{A}}_1}$} except ${{X}_1}$ (note that $X_1X_1={\bf{I}}$). The resulting state of Eq. \ref{eq-initialize-A1} is the target ground state corresponding to $\left\langle {{{\bf{A}}_1}} \right\rangle  = 1$. This process is shown in Fig. \ref{fig-representative-qubits}\textbf{A}. The key point to understand this process is that the representative qubit on the state $\left|  +  \right\rangle $ does not change under the action of ${{\bf{A}}_1}$. Thus, we can use it to implement ${{1 \over {\sqrt 2 }}\left( {{\bf{I}} + {\bf{A_k}}} \right)}$, {which is a superposition of the identity operator and string operator ${\bf{A_k}}$.} For a toric code with no {twist}, every qubit (except those on the edge) is shared by four square plaquettes. Thus, stabilizers corresponding to every plaquette can be assigned to an isolated representative qubit.

Since an entangled qubit cannot be stabilized by a single Pauli operator, the representative qubit should not be entangled to the rest qubits until it has controlled the corresponding ${{\bf{A_k}}}$. Because all ${{\bf{A_k}}}$ commute with each other, the operation ${{1 \over {\sqrt 2 }}\left( {{\bf{I}} + {\bf{A_k}}} \right)}$ will preserve stabilizers that are already well-prepared when the expected value of ${{\bf{A_k}}}$ is zero. For a state satisfying $\left\langle \phi  \right|{\bf{A_j}}\left| \phi  \right\rangle  = 1$ ($\bf j \ne k$), we have
\begin{align}\label{eq-no-interfer}
	\left\langle \phi  \right|{1 \over {\sqrt 2 }}\left( {{\bf{I}} + {{\bf{A}}_{\bf{k}}}} \right){\bf{A_j}}{1 \over {\sqrt 2 }}\left( {{\bf{I}} + {{\bf{A}}_{\bf{k}}}} \right)\left| \phi  \right\rangle  = \left\langle \phi  \right|\left( {{\bf{I}} + {{\bf{A}}_{\bf{k}}}} \right){\bf{A_j}}\left| \phi  \right\rangle  = 1 + \left\langle \phi  \right|{\bf{A_k}}\left| \phi  \right\rangle  = 1.
\end{align}
To keep the representative qubit not entangled with other qubits until all operations controlled by it are performed, we can activate the representative qubits layer by layer. The number of layers in the initialization circuit is proportional to the diameter of the rectangular system. This scaling is essentially optimal for the topologically ordered system \cite{Bravyi200607PRL}. We show a sketch of initializing the ground state of the toric code through the activating sequence of representative qubits in Fig. \ref{fig-representative-qubits}\textbf{B}.

\begin{figure}[htb]
	\includegraphics[width=1\linewidth]{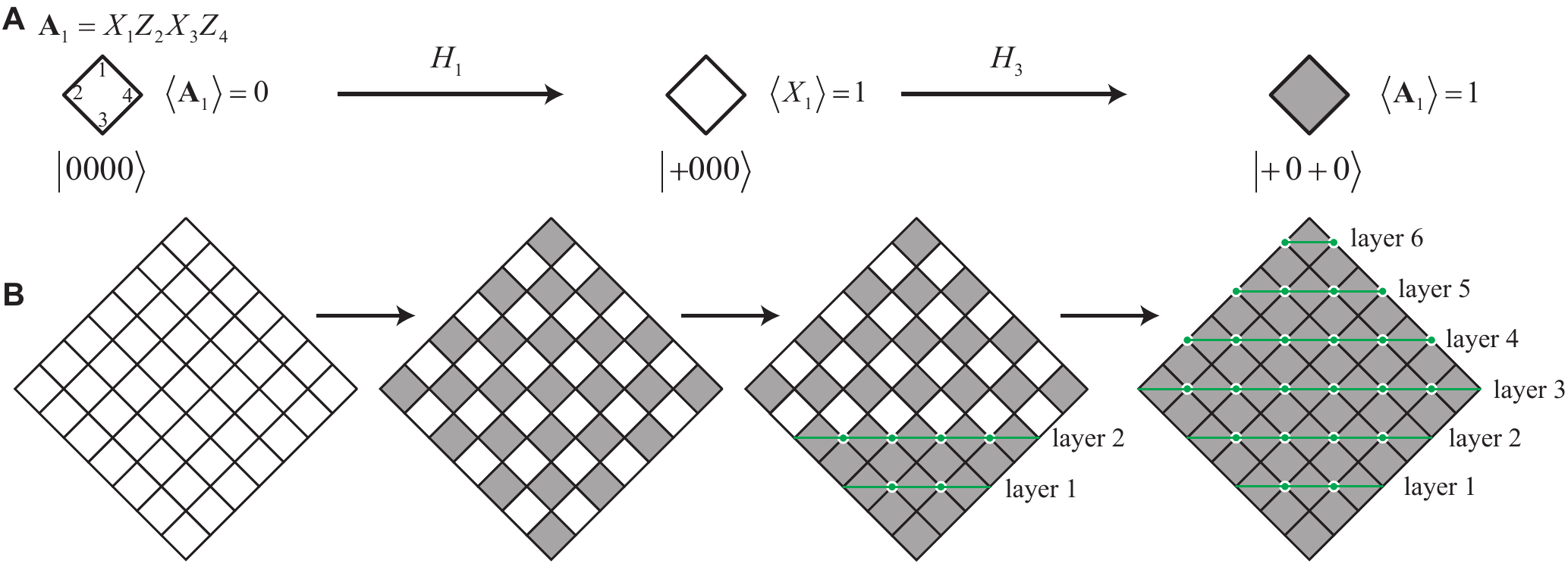}
	\caption{\label{fig-optimized-representative-qubits} The optimized method to prepare the ground state of the toric code. (\textbf{A})  The method to prepare the ground state of one independent plaquette, where the corresponding stabilizer is ${\bf{A}}_1 = {X_1}{Z_2}{X_3}{Z_4}$. This process can be described as ${{H}_1}{{H}_3}\left| {0000} \right\rangle  = \left| { + 0 + 0} \right\rangle $ and $\left\langle { + 0 + 0} \right|{\bf{A}}_1\left| { + 0 + 0} \right\rangle  = 1$. (\textbf{B})  The optimized method to prepare the ground state of the toric code. Under this scheme, half of the plaquettes can be prepared to the corresponding ground states by one layer of Hadamard gates. Thus, the number of representative qubits and the corresponding two qubits gates is halved.}
\end{figure}

There is a practical method to approximately halve the number of two qubits gates when initializing the ground state of toric code. Note that some stabilizers have no ``conflict'', which means they do not share common qubits attached by different Pauli operators. In other words, the ``conflict'' is where two different string operators cross. The tangent points are not regarded as ``conflicts'' since string operators have the same orientation at these points. According to the definition of string operators, the Pauli matrix attached to the tangent point of different string operators is the same. We only need to implement two Hadamard gates on the qubits attached to ${X}$ to obtain the ground state of an independent stabilizer ${\bf{A}}$, as shown in Fig. \ref{fig-optimized-representative-qubits}\textbf{A}. About half of the stabilizers can be initialized with this method, {where the corresponding plaquettes are marked in gray in the second figure of Fig. \ref{fig-optimized-representative-qubits}\textbf{B}.} These stabilizers have no conflict and can be prepared simultaneously. This method reduces the workload of representative qubits by half, and so does the number of layers. This optimized initialization method is shown in Fig. \ref{fig-optimized-representative-qubits}\textbf{B}.

\begin{figure}[htb]
	\includegraphics[width=1\linewidth]{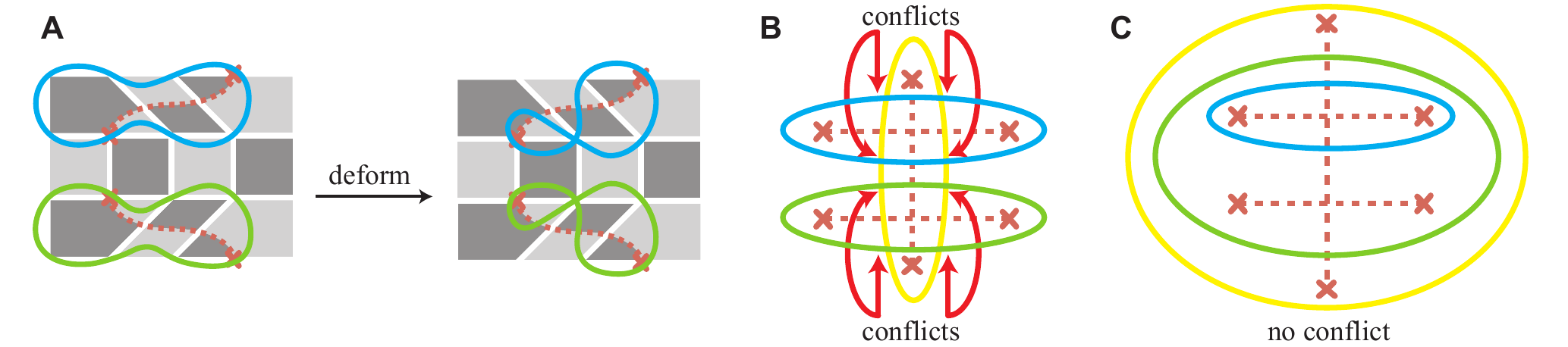}
	\caption{\label{fig-fermion-20}The string operators corresponding to the fermion charge. (\textbf{A}) Deforming string operators characterizing the fusion results of twists. The blue strings represent the fermion charge of the upper two twists and the green ones represent the fermion charge of the lower two. {The string operators on the right are the shortest ones containing the corresponding pairs of twists, respectively.} (\textbf{B})  The string operators that uniquely determine the fusion results of three pairs of twists. The eigenvalue of these string operators being $+1$ indicates the vacuum and being $-1$ indicates $\epsilon$. The desired fusion results corresponding to these string operators are recorded as $f_b$, $f_g$, and $f_y$ for the blue, green, and yellow string operators, respectively. These string operators have eight ``conflicts'' marked by red arrows, which make the method in Fig. \ref{fig-optimized-representative-qubits}\textbf{A} incapable to initialize them. (\textbf{C})  Three ``conflict''-free string operators uniquely determining the fusion results of three pairs of twists. The blue, green, and yellow string operators correspond to ${f_b}$, ${f_b} \times {f_g}$, ${f_b} \times {f_g} \times {f_y}$, respectively.}
\end{figure}

The existence of twists seems to reduce the number of terms in the system Hamiltonian, and reduce the difficulty of initialization. However, this is not true if we consider them as Ising anyons generated in pairs from the vacuum or fermion. A pair of twists will reduce one term in the system Hamiltonian. {This emergent degree of freedom is used to characterize the fusion result of this pair of twists. When we create a pair of twists from the vacuum (fermion), the string operator corresponding to the fermion charge of these two twists should be initialized to $ + 1\left( { - 1} \right)$.} Some topological equivalent string operators corresponding to the fermion charge of two pairs of twists on a $20$ qubits lattice are explicitly shown in Fig. \ref{fig-fermion-20}\textbf{A}. The initialization of string operators corresponding to fermion charge can be easily done based on the optimized initialization method shown in Fig. \ref{fig-optimized-representative-qubits}. One can firstly prepare the eigenstate of string operators that uniquely determine the fusion results of Ising anyons with the method of Fig. \ref{fig-optimized-representative-qubits}\textbf{A}. This method uses one layer of Pauli operators to initialize the system to the target state. Thus, the length of these string operators, or the distance between twists, does not influence {or slightly increase} the circuit depth. Then the rest stabilizers ${\bf{A_k}}$ can be initialized with the method shown in Fig. \ref{fig-representative-qubits}\textbf{B}. 

We note that the method shown in Fig. \ref{fig-optimized-representative-qubits}\textbf{A} requires that the string operators to be initialized have no ``conflict''. This method seems to fail in the case of many twists where the string operators characterizing the fusion results of twists have conflicts, as shown in Fig. \ref{fig-fermion-20}\textbf{B}. However, there is a practical trick to define string operators that uniquely determine the fusion results of twist and have no ``conflict''. As shown in Fig. \ref{fig-fermion-20}\textbf{B}, three string operators uniquely determine the fusion results of three pairs of twists. These results are recorded as $f_b$, $f_g$, and $f_y$ for the blue, green, and yellow string operators, respectively. String operators corresponding to these fusion results have eight ``conflicts'' and cannot be initialized with the method of Fig. \ref{fig-optimized-representative-qubits}\textbf{A}. This obstacle can be circumvented if we turn to initialize the string operators corresponding to $f_b$, ${f_b} \times {f_g}$, ${f_b} \times {f_g} \times {f_y}$, as shown in Fig. \ref{fig-fermion-20}\textbf{C}. These string operators have no ``conflict'' and can uniquely determine the fusion results of these three pairs of twists.

\subsection{Quantum circuit for braiding}

In topological quantum computing, the logic gate defined in the fusion space of twists can be implemented by braiding them. Here we analyze the quantum circuit to realize a single braiding of twists. One way to construct such a circuit is to calculate the square root of logic ${\bar Z}$ gate, since $\bar Z = B_{12}^2$. The unitary gate to implement logic ${\bar Z}$ is ${U_{\bar Z}} =  - i{c_1}{c_2}$, as introduced in the main text. We can calculate the unitary gate corresponding to a single braiding as
\begin{align}\label{eq-S-unitary}
	{U_{\bar S }} = \sqrt { -i{c_1}{c_2}}  = \sqrt { - i{e^{{\pi  \over 2}{c_1}{c_2}}}}  = {e^{ - {\pi  \over 4}i}}{e^{{\pi  \over 4}{c_1}{c_2}}} = {e^{ - {\pi  \over 4}i}}{1 \over {\sqrt 2 }}\left( {{\bf{I}} + {c_1}{c_2}} \right).
\end{align}
This is a non-local unitary and is hard to implement as a quantum circuit on large-scale quantum processors.

An alternative way to obtain the quantum circuit corresponding to a single braiding is to analyze the unitary gate that directly moves twists on the stabilizer graph. This process is to geometrically deform the stabilizer code so that the twists, combined with stabilizers characterizing them, move to the desired positions \cite{Bombin2010Topological, Bombin2009Quantum}. Without loss of generality, we consider one pair of different stabilizers between the original stabilizer graph and the one after moving the twist, denoted as ${S_{\text{old}}}$ and ${S_{\text{new}}}$, respectively. A simple way to achieve this deformation is to measure the new stabilizer ${S_{\text {new}}}$. However, this non-destructive measurement method is difficult to realize experimentally. The initialization methods shown in Figs. \ref{fig-representative-qubits} and \ref{fig-optimized-representative-qubits} also do not apply to the present situation because of two differences. The first one is that there is no free qubit not entangled with the system. Thus, the scheme of representative qubits fails. The second difference is that there will be one old stabilizer ${S_{\text {old}}}$ becomes invalid once the new stabilizer ${S_{\text {new}}}$ is activated. 

Now we explore a unitary gate that can change the stabilizer from $S_{\text {old}}$ to $S_{\text {new}}$. We note that these two stabilizers are Hermitian Pauli strings, their product multiplied by $i$ is Hermitian, written as:
\begin{align}\label{eq-Hermitian-S-product}
	{\left[ {i{S_{\text {new}}}{S_{\text {old}}}} \right]^\dag } =  - iS_{\text {old}}^\dag S_{\text {new}}^\dag  =  - i{S_{\text {old}}}{S_{\text {new}}} = i{S_{\text {new}}}{S_{\text {old}}}.
\end{align}
This is an important property since it means the operator
\begin{align}\label{eq-unitary-move-twist}
	{1 \over {\sqrt 2 }}\left( {{\bf{I}} + {S_{\text {new}}}{S_{\text {old}}}} \right) = {1 \over {\sqrt 2 }}\left[ {{\bf{I}} + i\left( { - i{S_{\text {new}}}{S_{\text {old}}}} \right)} \right] = {e^{{\pi  \over 4}i\left( { - i{S_{\text {new}}}{S_{\text {old}}}} \right)}} = {e^{{\pi  \over 4}{S_{\text {new}}}{S_{\text {old}}}}}
\end{align}
is a unitary operator. Note that ${{S_{\text {old}}}}$ is a stabilizer of the state $\left| \psi  \right\rangle $ before moving the twist. We can obtain
\begin{align}\label{eq-effect-move-twist}
	{1 \over {\sqrt 2 }}\left( {{\bf{I}} + {S_{\text {new}}}{S_{\text {old}}}} \right)\left| \psi  \right\rangle  = {1 \over {\sqrt 2 }}\left( {{\bf{I}} + {S_{\text {new}}}} \right)\left| \psi  \right\rangle .
\end{align}
Now we have the unitary operator ${e^{{\pi  \over 4}{S_{\text {new}}}{S_{\text {old}}}}}$ to achieve same effect of ${1 \over {\sqrt 2 }}\left( {{\bf{I}} + {S_{\text {new}}}} \right)$ on the state $\left| \psi  \right\rangle $ before moving the twist. Based on what we have discussed around Eq. \ref{eq-Hadamard-test-result}, the unitary operator ${e^{{\pi  \over 4}{S_{\text {new}}}{S_{\text {old}}}}}$ is equivalent to projecting $\left| \psi  \right\rangle $ to the eigenstate of ${{S_{\text {new}}}}$ with the corresponding eigenvalue equaling $ + 1$. In this way, we obtain a unitary gate to change the ``geometry'' of the stabilizer graph. The quantum circuit that sequentially implements such unitary gates can move and braid twists. Taking Fig. \ref{fig-code-deformation-30}\textbf{B} as an example, in the first step of moving the twist ${\sigma _1}$, ${S_{{\text{old}}}} = {Y_2}{Y_3}{Z_8}{X_9}$ and ${S_{{\text{new}}}} = {Y_2}{X_3}{Z_8}$. The corresponding unitary reduces to ${e^{i{\pi  \over 4}{Z_3}{X_9}}}$, which is a two-qubit gate. In the same way, we perform all the code deformations, which implement the desired braidings of twists in Fig. \ref{fig-code-deformation-30}, through a sequence of two-qubit gates. This figure shows the process of braiding two twists from different pairs and producing a pair of fermion charges step by step.

\begin{figure}[tb]
	\includegraphics[width=1\linewidth]{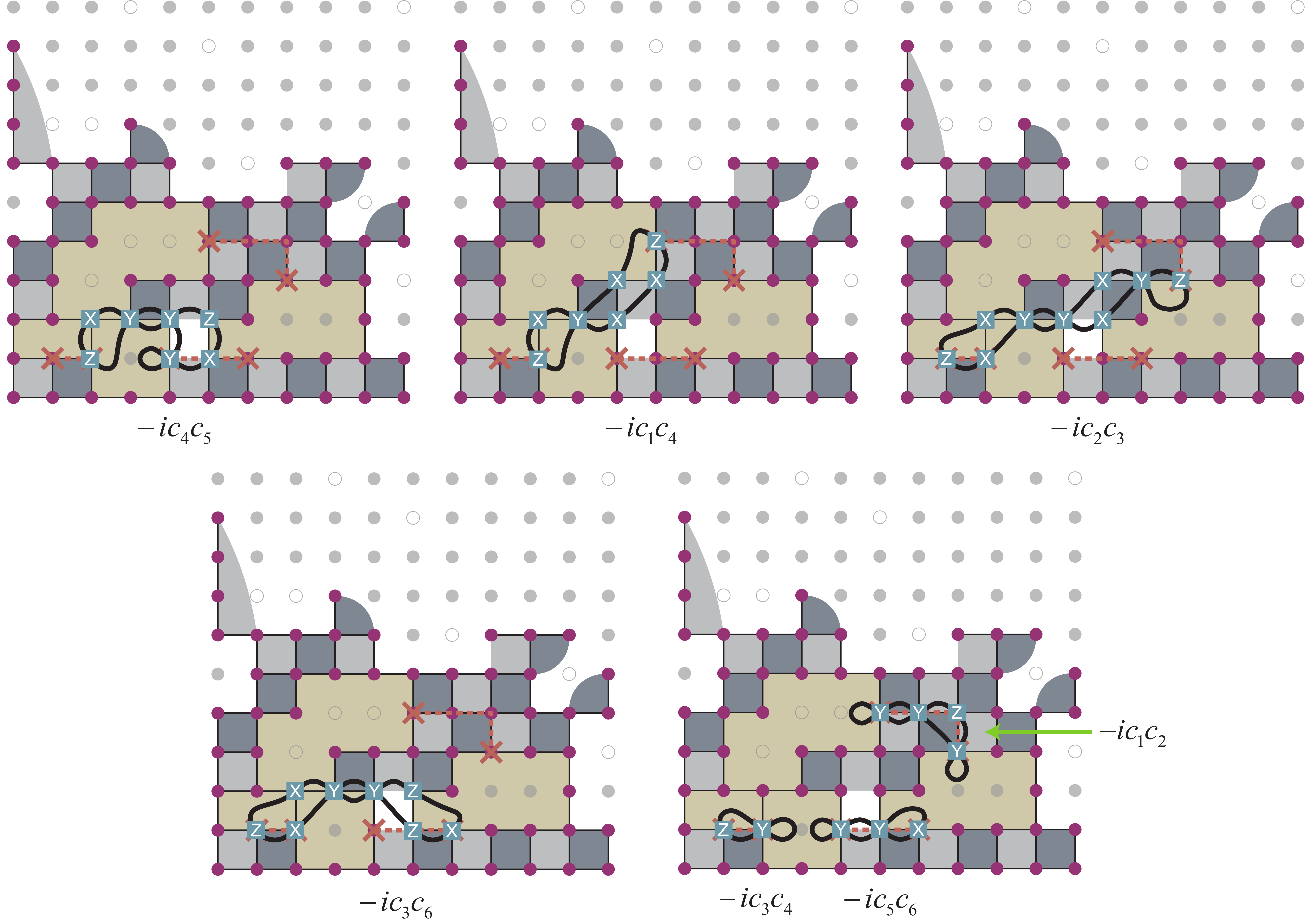}
	\caption{\label{fig-circuits-68}String operators and their attached Pauli matrices corresponding to $ - i{c_1}{c_2}$, $ - i{c_3}{c_4}$, $ - i{c_5}{c_6}$, $ - i{c_2}{c_3}$, $ - i{c_1}{c_4}$, $ - i{c_4}{c_5}$, and $ - i{c_3}{c_6}$. The subscripts of Majorana operators indicate the {indices of twists} being surrounded. For example, $ - i{c_1}{c_2}$ represents a string operator surrounding ${\sigma _1}$ and ${\sigma _2}$. It specifies the fusion result of these two {twists}. The complex phase factor is introduced to keep these operators both Hermitian and unitary. These two properties make them severing as quantum circuits and measurements performed on the physical qubit. Their implementations and corresponding logical operators are shown in Figs. 3 and 4 of the main text.}
\end{figure}

\subsection{Quantum circuit for processor I}\label{sec-circuit_for_processor_I}

In this subsection, we show the quantum circuits corresponding to the string operators that characterize the fusion results of Ising anyons in Fig. \ref{fig-circuits-68}. These string operators are both Hermitian and unitary, which are used as observables to be measured and quantum circuits to be implemented in Figs. 3 and 4 of the main text. Detailed introductions of these circuits are in the caption of Fig. \ref{fig-circuits-68}.

\section{Experimental Details}
\subsection{Device information}

\begin{figure}[htb]
	\centering\includegraphics[width=1\linewidth]{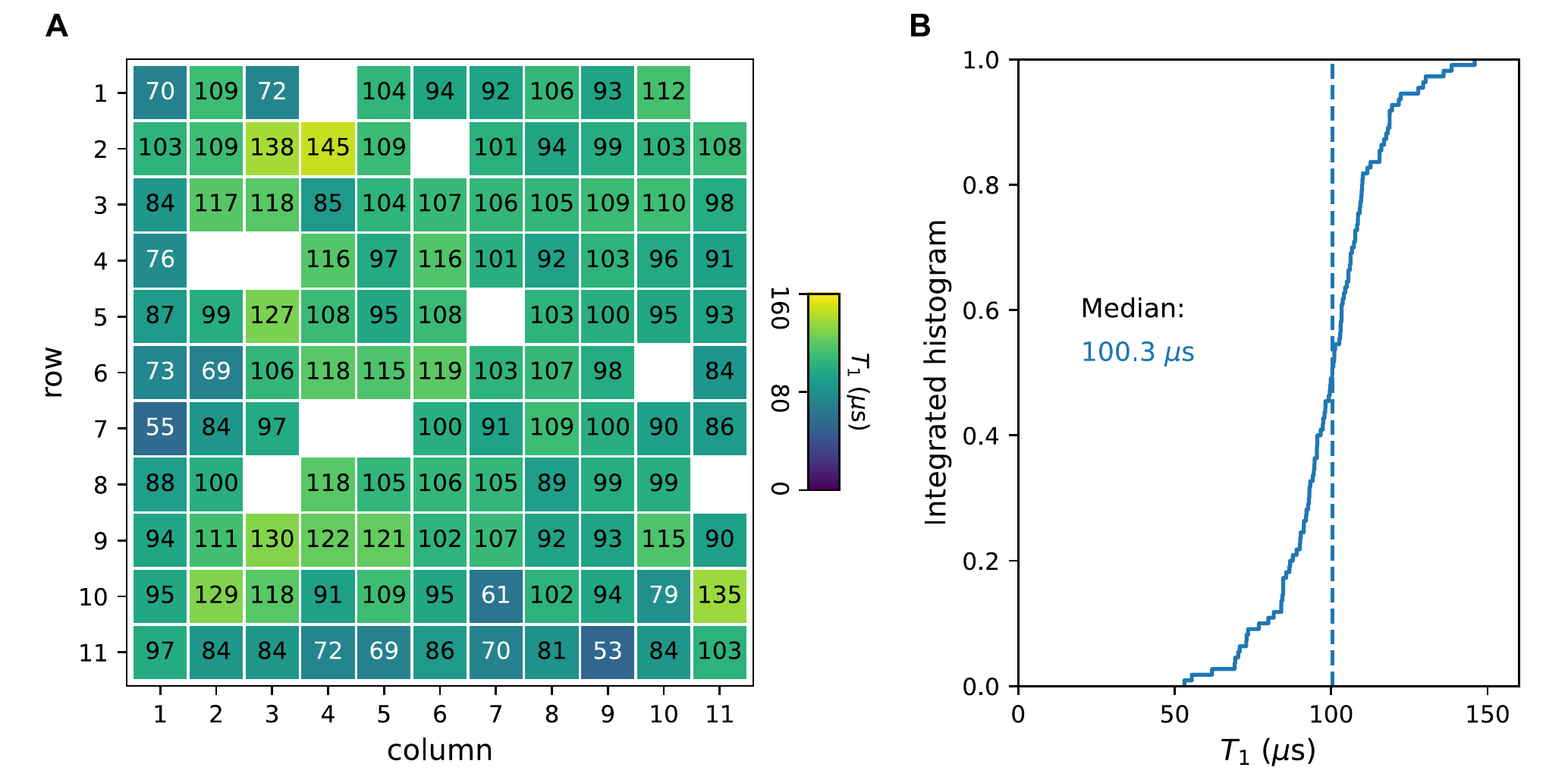}
	\caption{\label{fig-110q_T1}\textbf{Typical $T_1$ values of the 110 functional qubits on the processor I.} (\textbf{A})  Distribution of the $T_1$ time. We obtained the heatmap during two rounds of cooldowns due to the limitation of wirings in the dilution refrigerator and measurement electronics. $T_1$ values may vary with frequency, drift over time and incur a sudden change at a different cooldown. (\textbf{B})  Integrated histogram of $T_1$. The dashed line indicates the median value, from which we obtain that $T_1$ has a median value of $100.3$ 
		$\mu$s.}
\end{figure}

\begin{figure}[htb]
	\centering\includegraphics[width=1\linewidth]{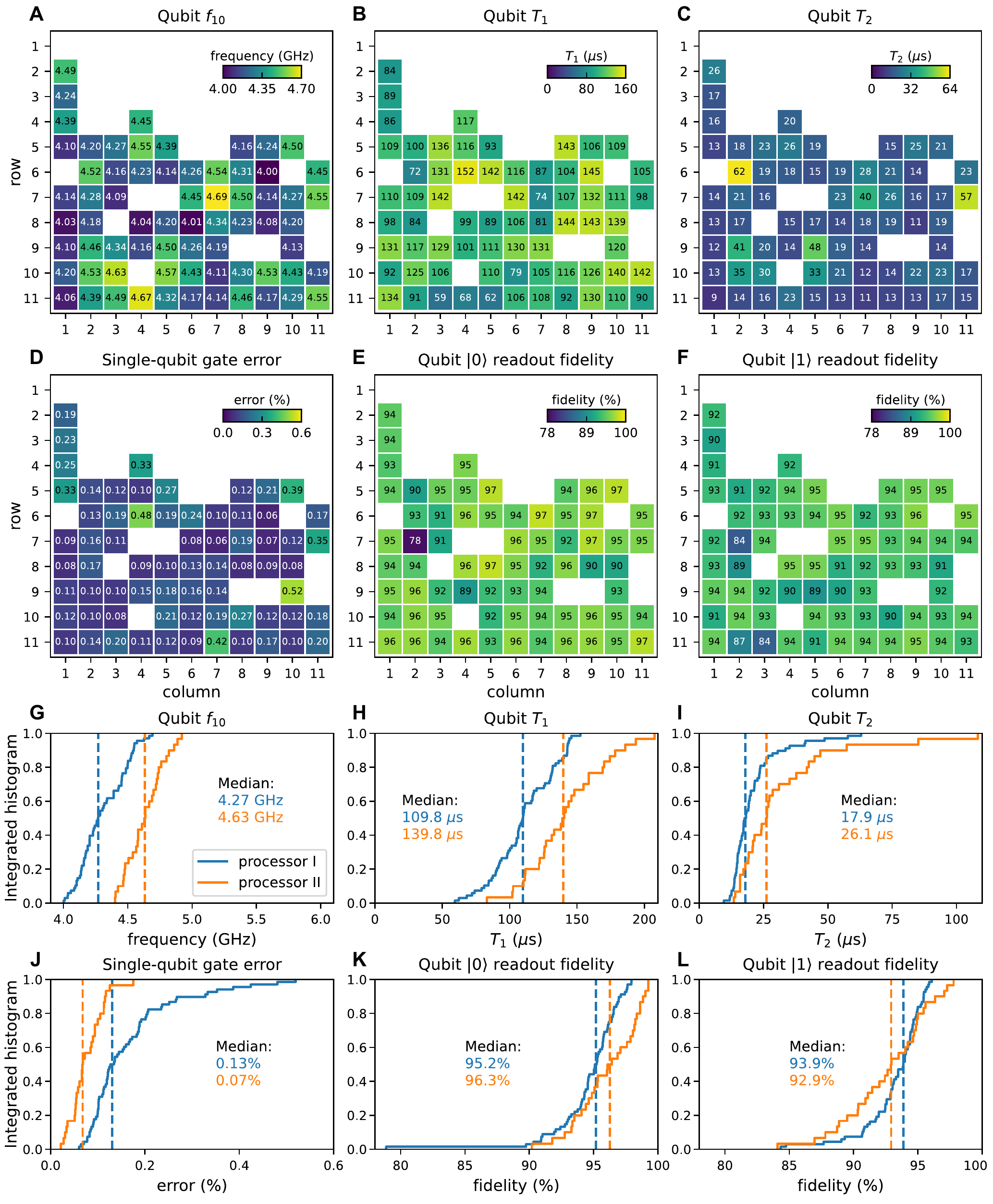}
	\caption{\label{fig-property_sq}\textbf{Heatmaps of various single-qubit parameters for processor I and corresponding integrated histograms for processors I and II.} (\textbf{A})  Qubit idle frequency. (\textbf{B})  Qubit relaxation time measured at the idle frequency. (\textbf{C})  Qubit dephasing time measured using Hahn echo sequence. (\textbf{D})  Pauli error of the single-qubit gate. (\textbf{E})  Readout fidelity of the qubit $|0\rangle$ state. (\textbf{F})  Readout fidelity of the qubit $|1\rangle$ state.  (\textbf{G-L}) Corresponding integrated histograms for processor I (blue) and II (orange). Dashed lines indicate the median values.}
\end{figure}

We observe the non-Abelian exchange statistics on two different quantum processors, referred to as %
processor I [Fig. 1(d) of the main text] and %
processor II \cite{processorII}, both of which were fabricated using the flip-chip recipe as described elsewhere \cite{Zhang2022Digital}. Processor I (II) hosts an array of $11\times11$ ($6\times6$) frequency-tunable transmon qubits with tunable couplers between adjacent qubits. On both processors, the maximum resonance frequencies of the qubit and coupler are around 4.8 GHz and 9.0 GHz, respectively. The effective coupling strength between two neighboring qubits can be dynamically tuned up to $-25$ MHz.
Each qubit capacitively couples to its own readout resonator, designed at the frequency around 6.5 GHz, for qubit state measurement.
{Processor I (II) possesses 110 (36) functional qubits, and the typical values of the energy relaxation times of these qubits are summarized in Fig. \ref{fig-110q_T1} (Ref. \cite{processorII}). Due to limited capacities in both wirings of our dilution refrigerator and measurement electronics, we measured all the qubits on processor I during two rounds of cooldowns,} 
and selected {a maximum number of 68 (30) qubits out of the 110 (36) functional qubits on processor I (II) for experiments {facing unexpected experimental realities such as broken wires resulting from thermal cycling}. The relaxation times and Hahn echo dephasing times measured at idle frequencies are shown in Fig. \ref{fig-property_sq}, with median values of $T_1 = 109.8 (139.8) \mu$s and $T_2 = 17.9 (26.1) \mu$s, respectively.} The cumulative distribution of the readout fidelities is also shown, which are used to mitigate measurement errors \cite{20q_ghz}. In addition, we also plot the distribution of single-qubit parameters such as idle frequencies and Pauli errors of single-qubit gate on processor I for the 68 qubits.

\begin{figure}[htb]
	\centering\includegraphics[width=1\linewidth]{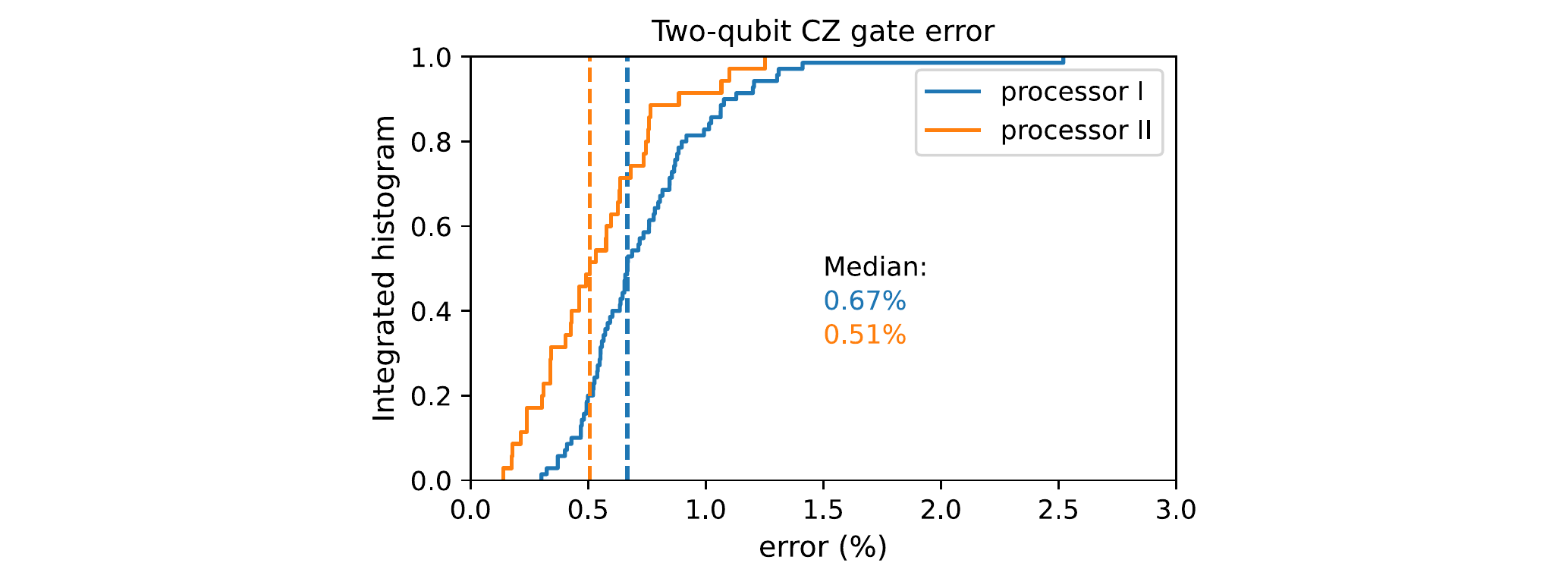}
	\caption{\label{fig-property_tq}{\textbf{Integrated histograms of the two-qubit CZ gate Pauli errors on processors I and II for 68 (30) qubits with the sample sizes of 70 and 35, respectively.}}
		Dashed lines indicate the median values.}
\end{figure}

\subsection{Calibration procedure}
Quantum operations on our superconducting quantum processors are physically realized by applying on each qubit/coupler analog signals with continuous control parameters such as amplitudes and phases.
A calibration procedure is a collection of experiments that is used to learn and optimize these control parameters and enables us to gain full control of the processor.
A systematic, automated and scalable calibration procedure is essential for achieving high-fidelity qubit operations across the whole device. 
We separate our calibration procedure into two sessions: a \textit{single-qubit calibration} session that is used to bring up all the qubits/couplers individually from scratch and collect basic device and control parameters, and a \textit{multiple-qubit calibration} session that is used to calibrate the processor on a system level and achieve high-fidelity simultaneous single- and two-qubit gates. 
Below we illustrate a couple of key steps during the bootstrapping calibration procedure.
\subsubsection{Single-qubit calibration}
In the single-qubit calibration session, we first bring up each qubit individually. We isolate each qubit from the others by applying on it a flux bias to set its transition frequency to around 4.5 GHz while keeping the other qubits/couplers at their sweet points. Then we bring up the qubit following the procedure as summarized below:
\begin{itemize}
	\item Perform qubit spectroscopy and power Rabi oscillations to optimize the parameters (drive frequency and power) of the $\pi$ pulse that excites the qubit to $|1\rangle$ state. 
	\item Perform qubit spectroscopy as a function of bias to find the relation between the qubit transition frequency and bias.
	\item Calibrate the timing between the qubit microwave pulse and flux bias pulse.
	\item Measure the response of the qubit to a detuning pulse to eliminate the impact of pulse distortions \cite{pulse_shape_dicarlo, ORBIT_Kelly}. 
	\item Measure the spectrum of $T_1$, which is later used in allocating the qubit energy level in the multiple-qubit calibration session.
\end{itemize}

After bringing up all the qubits, we move on to bring up all the couplers. We isolate each coupler from the others by applying on it a flux bias to set its transition frequency to around 5.6 GHz while keeping the other couplers at their sweet points. At the same time, we set one of its nerghboring qubits to around 4.5 GHz, which is used to readout the coupler state, and another to below 4 GHz, while keeping other qubits at their sweet points. Then we bring up the coupler following the procedure as summarized below:
\begin{itemize}
	\item Calibrate the timing between the microwave pulse of the readout qubit and flux bias pulse of the coupler by applying a square flux bias pulse on the coupler to dispersively detune the qubit.
	\item Measure the swapping dynamics of the readout qubit and the coupler by applying a flux bias pulse on the coupler with a fixed time of 5 ns and vary its amplitude. Determine the amplitude with which a complete state transfer between the coupler and readout qubit is realized. 
	\item Perform coupler spectroscopy and power Rabi oscillations to optimize the parameters (drive frequency and power) of the $\pi$ pulse that excites the coupler to $|1\rangle$ state. The microwave pulse is applied via the microwave line of the readout qubit.
	\item Measure the response of the coupler to a detuning pulse to eliminate the impact of pulse distortions.
\end{itemize}

\begin{figure}[htb]
	\includegraphics[width=1\linewidth]{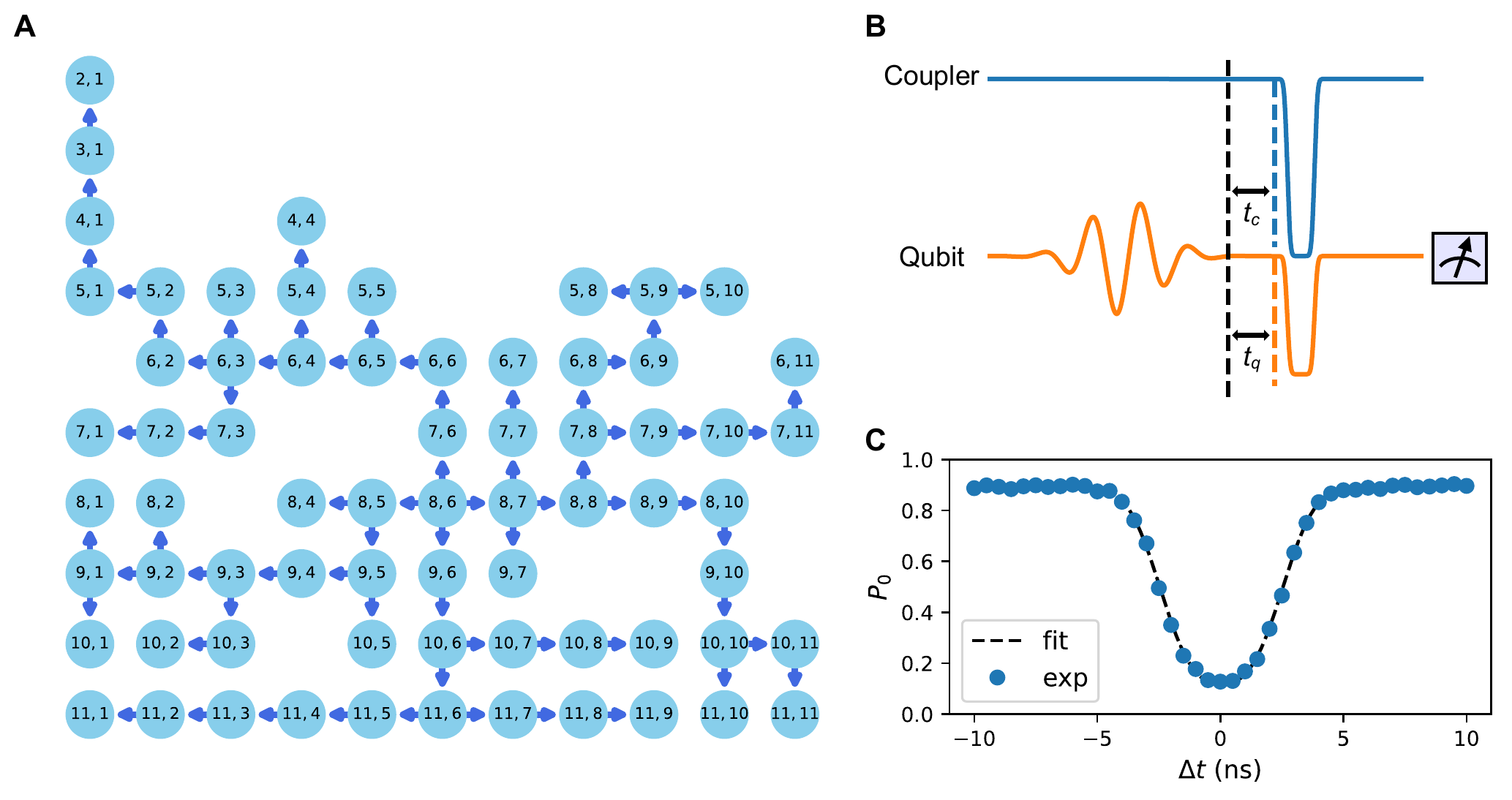}
	\caption{\label{fig-timing_tree} \textbf{Synchronization of the flux bias pulses of qubit-coupler pairs.} (\textbf{A})  The directed acyclic graph (DAG) of the 68-qubit system is constructed with the breadth-first search (BFS) algorithm, which is used to align the timing among all the flux bias pulses across the device. Each circle denotes a qubit indexed by the numbers inside. The calibration is performed in the order indicated by the direction of the arrows. (\textbf{B})  Experimental pulse sequences for calibrating the timing between the flux bias pulses of a qubit-coupler pair. After exciting the qubit to $|1\rangle$ with a $\pi$ pulse and waiting for a fixed delay $t_c$, we apply a flux bias pulse with a length of around 5 ns on the coupler, which can transfer the photon from the qubit to the coupler when the qubit stays at the idle frequency. At the same time we apply a flux bias pulse with the same length on the qubit and adjust its beginning time $t_q$. (\textbf{C})  The experimentally measured $|0\rangle$-state probability (dots) of the qubit vs. the time offset $\Delta t (=t_q-t_c)$. When the two flux bias pulses are synchronized, the qubit will stay at $|1\rangle$ state, resulting in a dip on the $P_0$ curve. We fit the location of the dip to adjust the timing between the two pulses.}
\end{figure}

Now that we have tuned up all the qubits and couplers individually, we move on to do calibrations that require a qubit-qubit or qubit-coupler pair at a time:
\begin{itemize}
	\item Construct a directed acyclic graph (DAG) according to the topological structure of the device with the breadth-first search (BFS) algorithm (Fig. \ref{fig-timing_tree}\textbf{A}), and fine tune the timing between the flux bias pulses of each qubit-coupler pair from the root to every branches in the DAG. The pulse sequences and typical data are shown in Fig. \ref{fig-timing_tree}\textbf{B}. After this step, all the control pulses are synchronized.
	\item Characterize the flux bias crosstalk matrix among all the qubits and couplers, which is used to actively cancel out the flux bias crosstalk effect \cite{20q_ghz}.
\end{itemize}

\subsubsection{Multiple-qubit calibration}
\begin{figure}[htb]
	\includegraphics[width=1\linewidth]{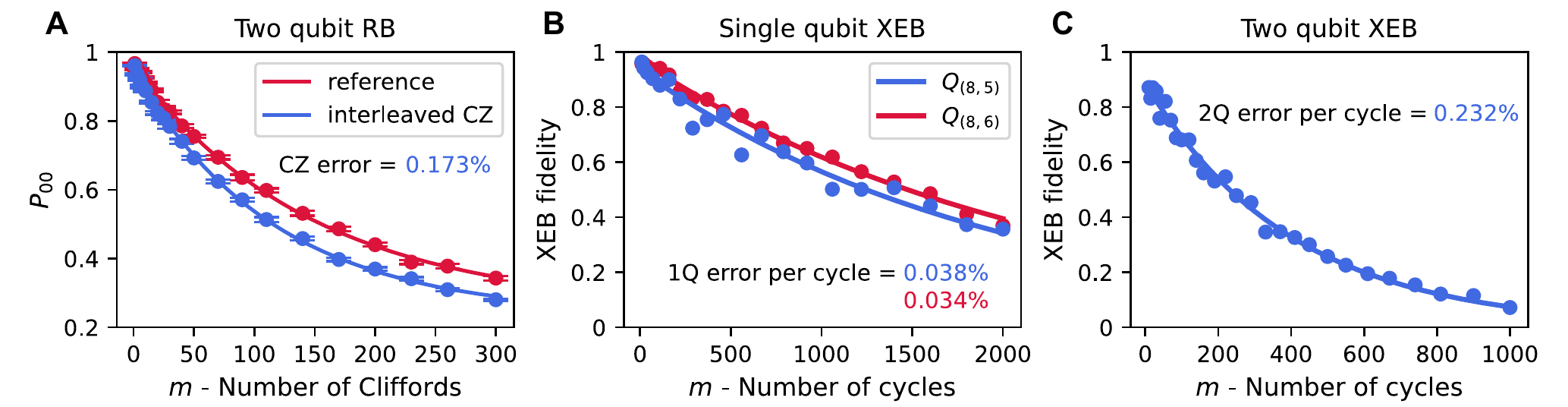}
	\caption{\label{fig-high_fidelity_cz} \textbf{Best individual CZ gate.}
		(\textbf{A}) Interleaved RB data
		with a sample size of $n=30$.
		(\textbf{B})  Simultaneous XEB data to characterize the single-qubit gates.
		(\textbf{C})  XEB data to characterize the same CZ gate as in (\textbf{A}), where each cycle contains two single-qubit gates in parallel and a CZ gate.
		The CZ Pauli errors extracted from RB and XEB are $0.173\%$ and $0.160\%$, respectively.}
\end{figure}

In the multiple-qubit calibration session, we allocate all the qubits and couplers to proper frequencies with the knowledge obtained from the previous session, based on which we can achieve high-fidelity single- and two-qubit gates. The procedure is summarized below:
\begin{itemize}
	\item Allocate the energy levels of all the qubits by solving a constraint optimization problem with a Satisfiability Modulo Theory (SMT) solver \cite{pysmt2015}. With the constraints we avoid the impact of various factors such as TLS defects, stray couplings between qubits, carrier frequency leakage, and pulse distortions.  We then allocate the readout energy levels of all the qubits in a similar way. %
	\item Characterize the microwave crosstalk coefficients for qubit pairs that are allocated within 60 MHz, which are used to actively cancel out the microwave crosstalk effect \cite{quantum_adversarial_learning}.
	\item Allocate the idle frequencies of all the couplers to minimize the unwanted qubit-qubit coupling. 
	\item Fine tune parameters (qubit resonant frequency, drive power, DRAG coefficient) to realize high-fidelity simultaneous single-qubit $\pi$ and $\pi/2$ pulses.
	\item Fine tune parameters (qubit bias, coupler bias, interaction frequency) to realize high-fidelity individual two-qubit CZ gates. We achieve a CZ gate with a fidelity as high as 0.9987 during this procedure (Fig. \ref{fig-high_fidelity_cz}).
	\item Benchmark the performance of the CZ gates when executed simultaneously for each two-qubit gate layer in the experimental circuit. Adjust the interaction frequencies when necessary.
\end{itemize}

\subsection{Experiment circuit}

\begin{figure}[htb]
	\centering
	\includegraphics[width=1\linewidth]{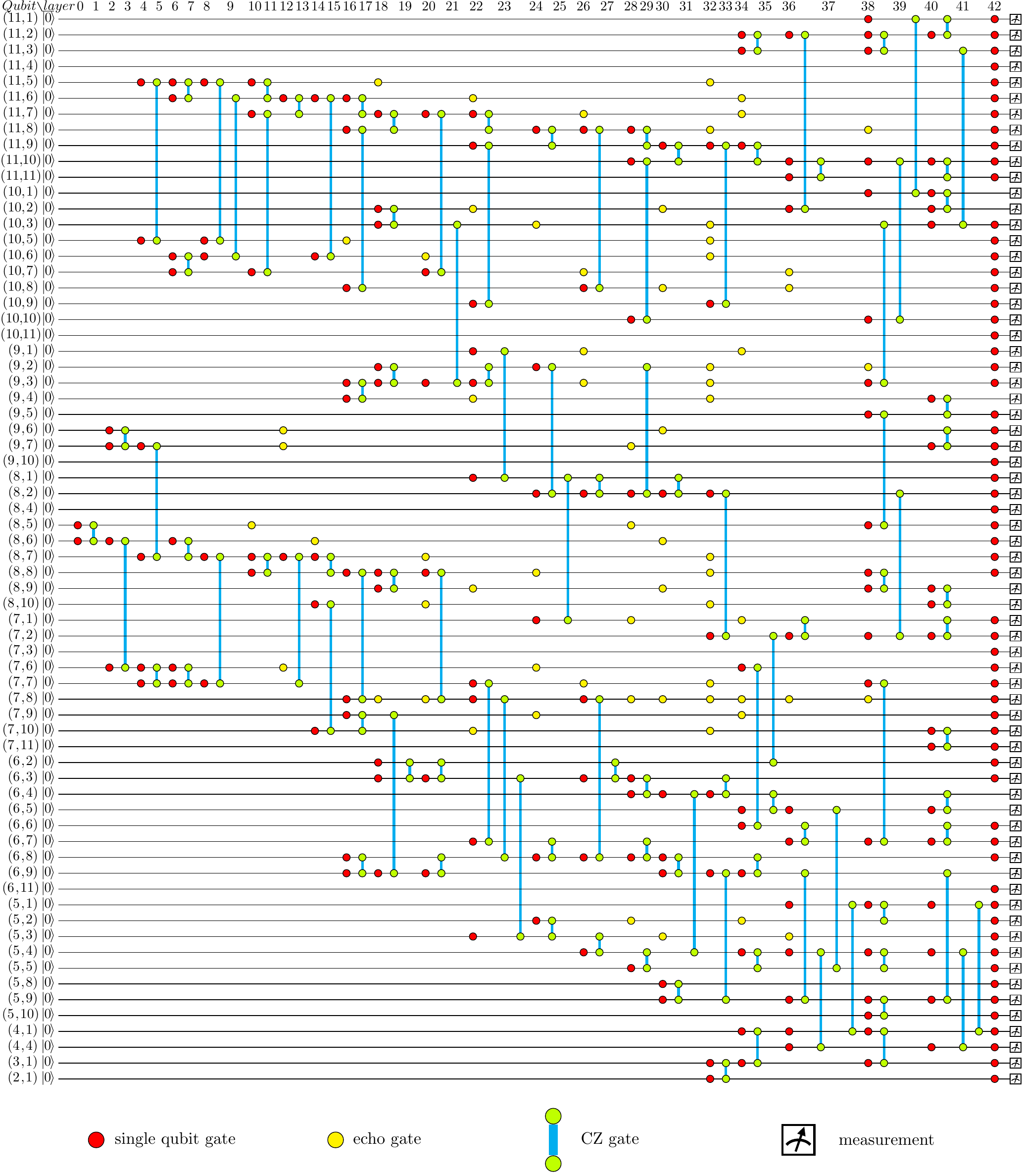}
	\caption{\label{fig-circuit-68q} \textbf{Experimental circuit for preparing the ground state with the 68-qubit system.} 
		The whole circuit contains 43 layers (labeled from 0 to 42 at the top of the figure), with each layer containing only single- or two-qubit gates that are applied simultaneously. Every single-qubit gate (red dots) is a Clifford gate realized by combining XY and Z rotations experimentally. The additional echo gates (yellow dots) are implemented by Y gates. There are in total 294 single-qubit gates (including 66 echo gates) and 113 two-qubit CZ gates.}\label{Fig-68-quantum circuits}
\end{figure}

\begin{figure}[htb]
	\includegraphics[width=0.5\linewidth]{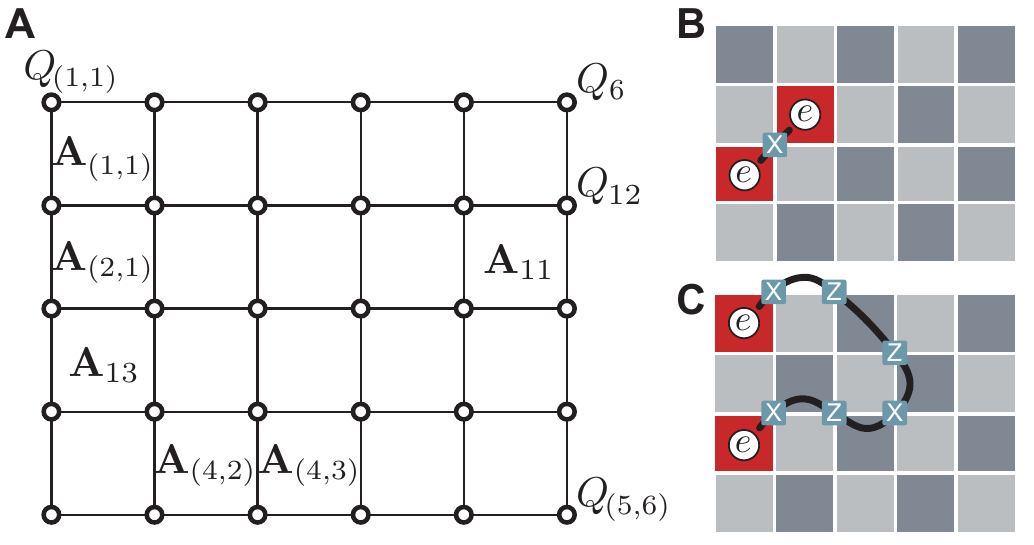}
	\caption{\label{fig-toric-code-30}{\textbf{The layout of the chosen 30 qubits on processor II and string operators.} (\textbf{A})  A sketch of the toric code lattice with 30 qubits on processor II. We index the qubits and stabilizers in two ways: one with two numbers corresponding to the row and column indices, and the other with the Zig-Zag index. 
			For example, $Q_{\left({2,6}\right)}$ and ${Q_{12}}$ denote the same qubit, and ${\bf{A}}_{\left( {3,1} \right)}$ and ${\bf{A}}_{13}$ denote the same plaquette operator. 
			(\textbf{B}) The creation of a pair of $e$ anyons with a single Pauli-X gate. The possible positions where charge $e$ and $m$ may arise are colored in dark and light grey, respectively. The red plaquettes indicate where the $e$ anyons live, with $\langle{\bf{A_k}}\rangle =  - 1$. (\textbf{C})  An example of a string operator that creates two $e$ anyons at its end, which can also be regarded as dragging one $e$ anyon from one end to the other.}}
\end{figure}

The circuit for preparing the ground state in the 68-qubit system is plotted in Fig. \ref{fig-circuit-68q}, which contains 22 layers of single-qubit gates and 21 layers of two-qubit CZ gates. 
In our experiment, arbitrary single-qubit gates can be realized by combining XY and Z rotations. 
{XY rotations are realized by applying 30-ns-long microwave pulses resonant with qubit frequency, with a full-width half-maximum of 15 ns and shaped with DRAG technique \cite{PhysRevLett.119.180511}, and Z rotations are realized with virtual-Z gate \cite{PhysRevA.96.022330}. }
During the application of XY pulses, we use active microwave cancellation technique at each qubit's frequency to mitigate the microwave crosstalk effects. The two-qubit CZ gate is realized by dynamically steering the resonant frequency of the coupler along a well-designed trajectory, so that the effective ZZ-type coupling strength can be turned on for a specific amount of time, accumulating a conditional $\pi$ phase angle while minimizing the leakage. CZ gate parameters are optimized for each two-qubit CZ gate layer. 
To characterize the performance of the quantum gates used in this experiment, we perform simultaneous cross entropy benchmarking (XEB).
{The median Pauli errors are $1.3\times10^{-3}$ ($7\times10^{-4}$)
	for the single-qubit gates and $6.7\times10^{-3}$ ($5.1\times10^{-3}$) for the two-qubit gates on Processor I (II), with the experimental data shown in Figs. \ref{fig-property_sq}\textbf{J} and \ref{fig-property_tq}.}

After constructing the original circuit for generating the ground state with the method depicted in \ref{sec_circuit_for_init}, we use ZX-calculus \cite{kissinger2020Pyzx} and Qiskit \cite{Qiskit} to reduce the depth of the circuit and recompile it with single- and two-qubit gates selected from the gateset \{RX, RY, RZ, CZ\}, where RX, RY and RZ denote single-qubit rotations around the x-, y-, and z-axis in the Bloch sphere, and CZ denotes the two-qubit CZ gate. Then we align all gates right while avoiding performing single- and two-qubit gates simultaneously. As a result, the optimized circuit consists of a staggered arrangement of single- and two-qubit gate layers. During the execution of the circuit, we apply CPMG sequences to minimize the impact of dephasing, which is realized by applying two Y gates (rotation around the y-axis by $\pi$ angle) on qubits whose successive idle layers span more than six single-qubit gate layers.

{
	\subsection{Error mitigation}
	Since the total topological charge is $\mathbf{1}$, $\epsilon$ fermions are expected to appear in pairs. As a result, post-selection upon an even number of fermions is legitimate to partially mitigate the error induced by gate imperfections and decoherence.
	In our experiment, ${-i{{c}}_{1}{{c}}_{2}}$, ${-i{{c}}_{3}{{c}}_{4}}$ and ${-i{{c}}_{5}{{c}}_{6}}$ can be sampled simultaneously. With the expectation values of these operations we can distinguish the parity of the number of fermions. Specifically, only zero or two out of the three expectation values are allowed to be -1.
	Here we have performed post-selection, whenever possible, for the experimental results shown in Figs. 3 and 4 of the main text.
}

\section{Additional Experimental results on 30 qubits}
\begin{figure}[htb]
	\includegraphics[width=0.5\linewidth]{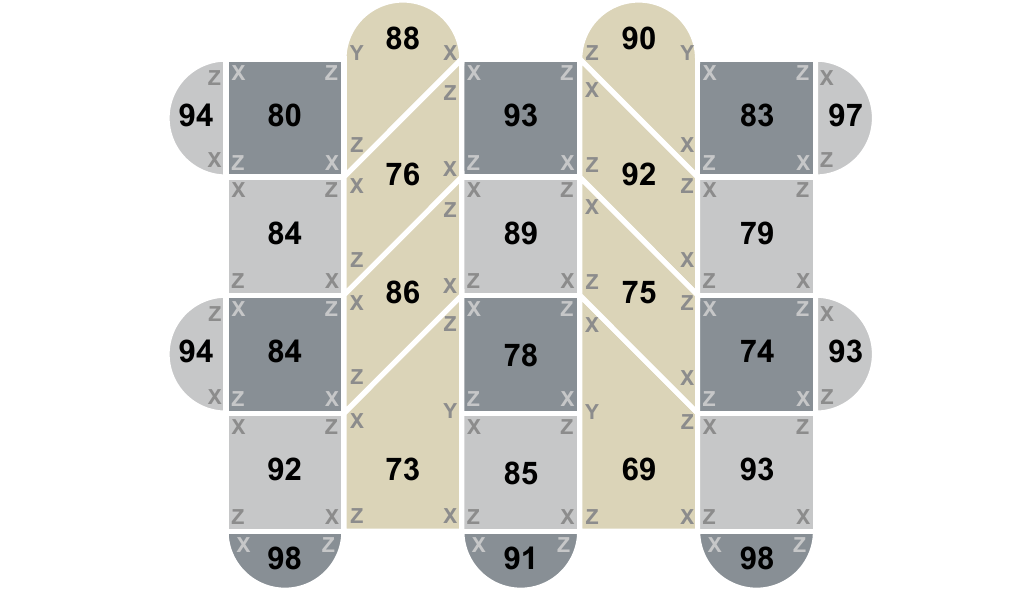}
	\caption{\label{fig-ground-state-30} \textbf{Experimental results of the ground state preparation on processor II with 30 qubits.} The color block marked with integer numbers surrounded by white edges represent stabilizers, which can be expressed as products of the Pauli operators. These Pauli operators are explicitly marked on the corresponding blocks. The integer numbers show the measured stabilizer values (in percentage) of the prepared deformed toric-code state. The average stabilizer value is 0.86. See also Fig. \ref{fig-code-deformation-30}\textbf{C}.
	}
\end{figure}

\begin{figure}[htb]
	\includegraphics[width=0.5\linewidth]{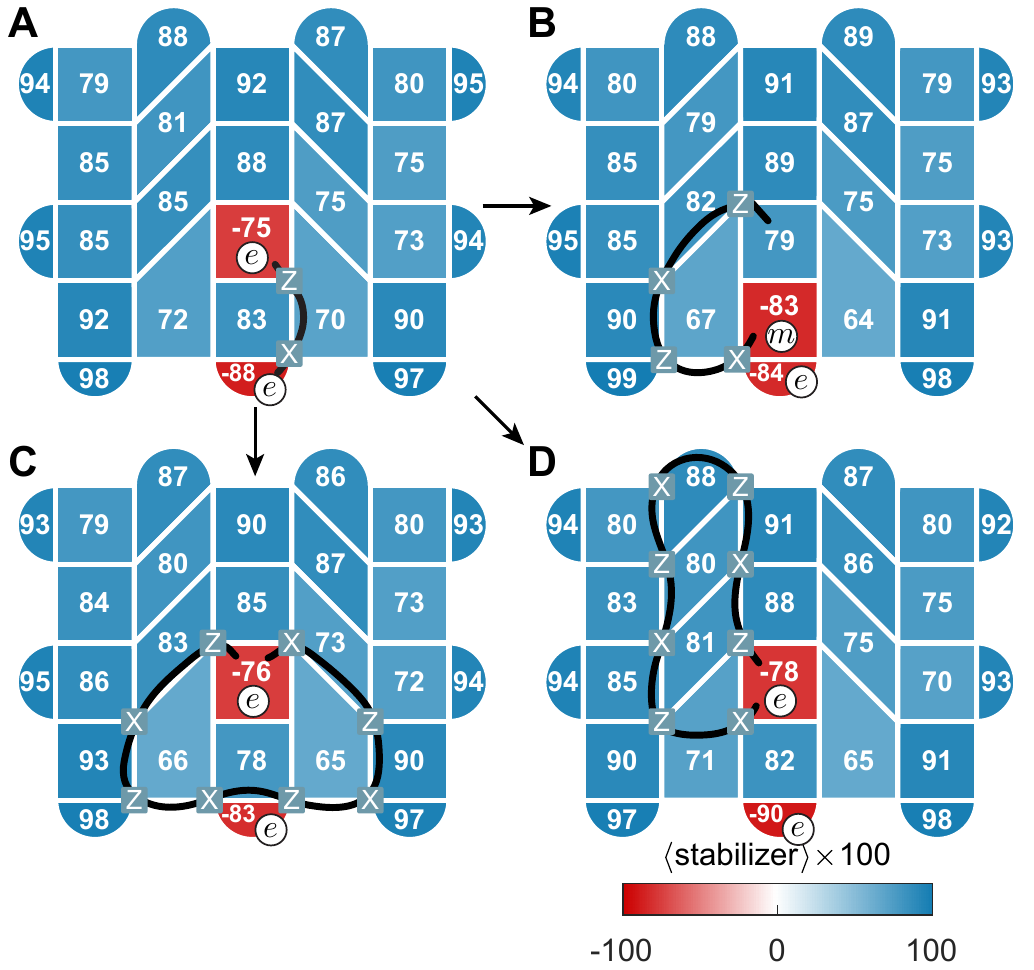}
	\caption{\label{fig-exchange-charge-30} \textbf{Experimental demonstration of exchanging charges between $e$ and $m$ on processor II with 30 qubits.} (\textbf{A}) The creation of a pair of $e$ anyons with a string operator. (\textbf{B}) The transition from an $e$ anyon to an $m$ anyon after cycling one twist. (\textbf{C-D}) The conservation of the $e$ anyon after cycling two twists generated either by different deformations or by the same deformation. We measure all stabilizers at each step, whose values (in percentage) are shown in the corresponding blocks. %
	}
\end{figure}

\begin{figure}[htb]
	\includegraphics[width=1\linewidth]{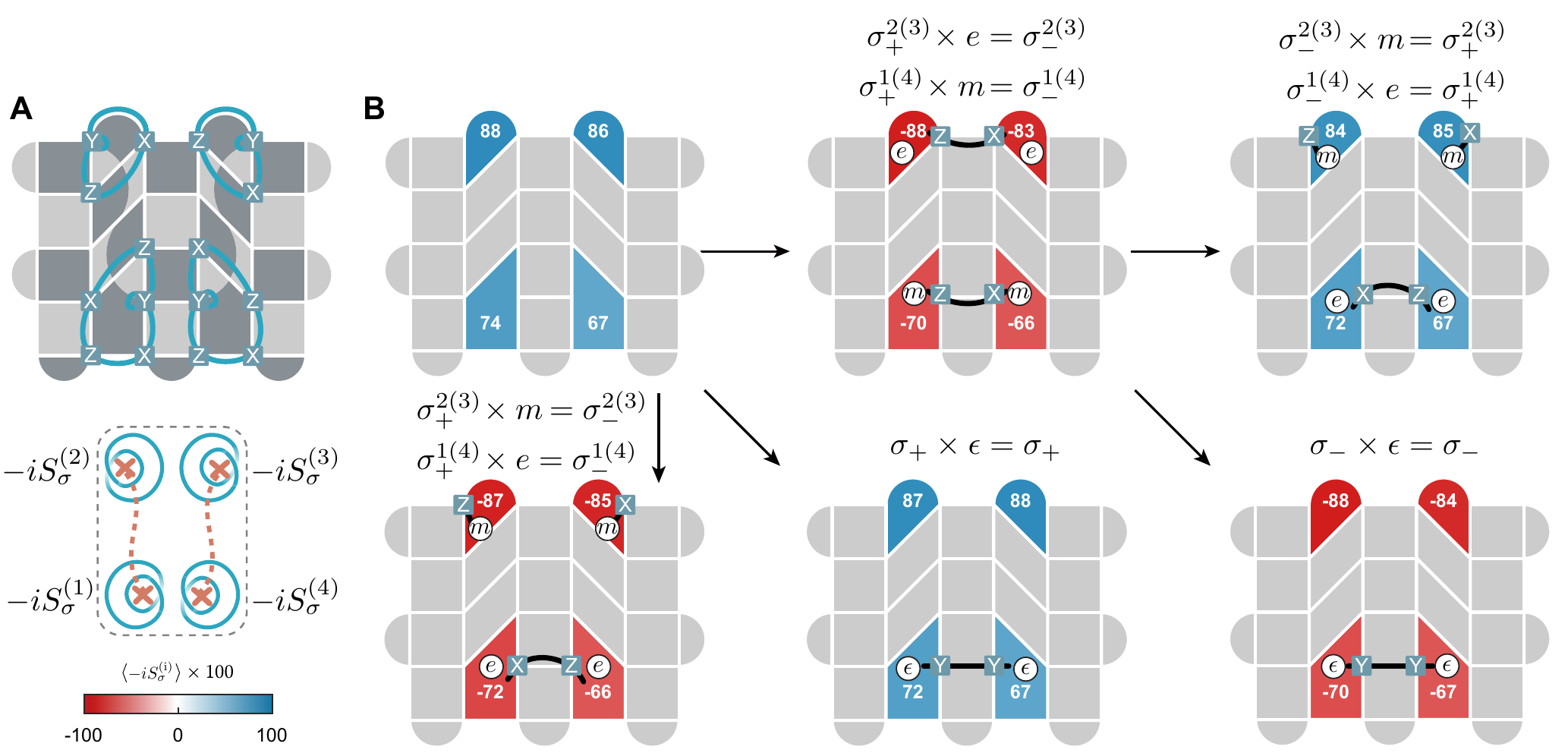}
	\caption{\label{fig-fusion-30} \textbf{Demonstration of fusion rules on processor II.} (\textbf{A})  The string operators $-iS_{\sigma}$ that specify the generalized charges of the four {twists}. A conceptual graph demonstrating the relative positions of the twists and the corresponding string operators is shown at the bottom. (\textbf{B})  Experimental results for verifying some fusion rules in Eq. 2 of the main text. We start from the ground state with four ${\sigma _ + }$ {twists} (blue blocks shown in the upper left panel). We verify the fusion rules between ${\sigma _ + }$ and $e$, $m$, $\epsilon$ by creating the corresponding quasiparticles in the deformations respectively. The lower left panel and the upper middle panel show that by fusing a ${\sigma _ + }$ {twist} with either an $e$ or an $m$ anyon we can obtain a $\sigma _-$ particle; The lower middle panel shows that by fusing a ${\sigma _ + }$ {twist} with an $\epsilon$ anyon we  obtain a ${\sigma _ + }$ particle again. In the right-most two panels, we use the ${\sigma _ - }$ {twists} (red blocks) generated at the previous step to verify the fusion rules between ${\sigma _ - }$ and $e$, $m$, $\epsilon$ quasiparticles, respectively. The integer numbers (in percentage) in the blocks show the measured stabilizer values.}
\end{figure}

\begin{figure}[htb]
	\includegraphics[width=1\linewidth]{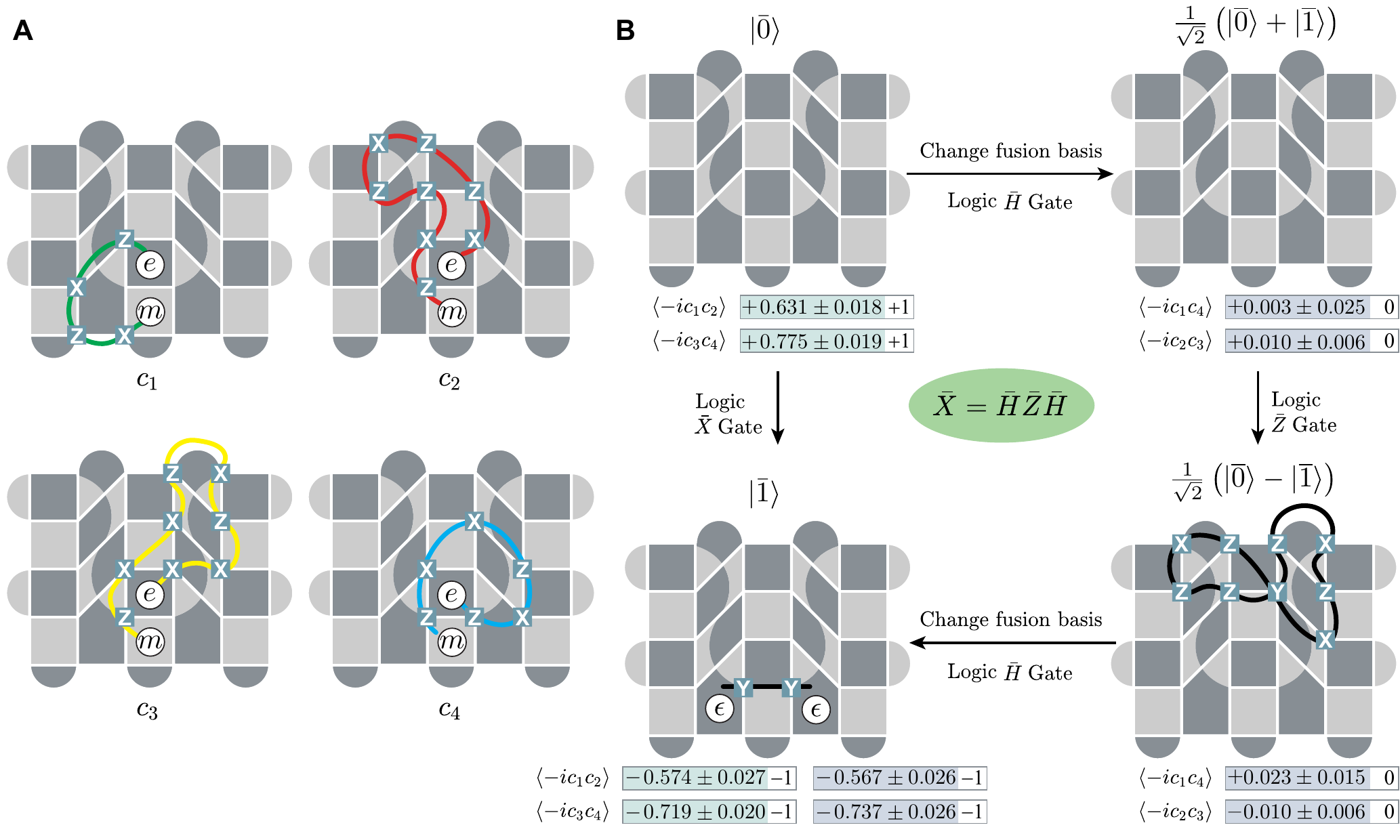}
	\caption{\label{fig-logical-gates-30} \textbf{The Majorana operators and demonstration of logic gates on processor II.} (\textbf{A})  Four Majorana operators corresponding to the four {twists}. (\textbf{B})  The demonstration of $\bar H \bar Z \bar H \left| {\bar 0 } \right\rangle  =\bar X  \left| {\bar 0} \right\rangle $ in the logical space, which is realized by braiding the twists described by the four Majorana operators. Here, the encoding of the logical qubit is $|\bar{0}\rangle=|(\sigma_1\times\sigma_2)_{\mathbf{1}},(\sigma_3\times\sigma_4)_{\mathbf{1}}\rangle$. The experimentally measured Majorana correlations (shaded values with error bars) and the corresponding ideal theoretical values (unshaded integers) are shown at the bottom of each panel at each step, which correspond to the expectation values of the logical $\bar{Z}$ observable. The first logic $\bar H $ gate is implemented by changing the Majorana operators as ${c_1} \mapsto {c_3}$, ${c_2} \mapsto {-c_2}$, ${c_3} \mapsto {c_1}$, and ${c_4} \mapsto {c_4}$. Thus, the fusion result of ${\sigma _1}$ and ${\sigma _2}$ changes from $ - i{c_1}{c_2}$ to $ - i{c_2}{c_3}$, and the fusion result of ${\sigma _3}$ and ${\sigma _4}$ changes from $ - i{c_3}{c_4}$ to $ - i{c_1}{c_4}$. The logical $\bar{Z}$ operator is applied on the changed fusion basis, which is $ - i{c_2}{c_3}$ instead of $ - i{c_1}{c_2}$. The second logic $\bar{H}$ gate recovers the indices and phase factors of these Majorana operators to their original values. On the other side of the equation, the logic $\bar{X}$ gate is implemented by the string operator $-i{c_1}{c_4}$ instead of the previously defined ${{{U}}_{\bar X}} =  - i{c_2}{c_3}$ based on the conservation of the total charge of the four twists.}
\end{figure}

\begin{figure}[htb]
	\includegraphics[width=1\linewidth]{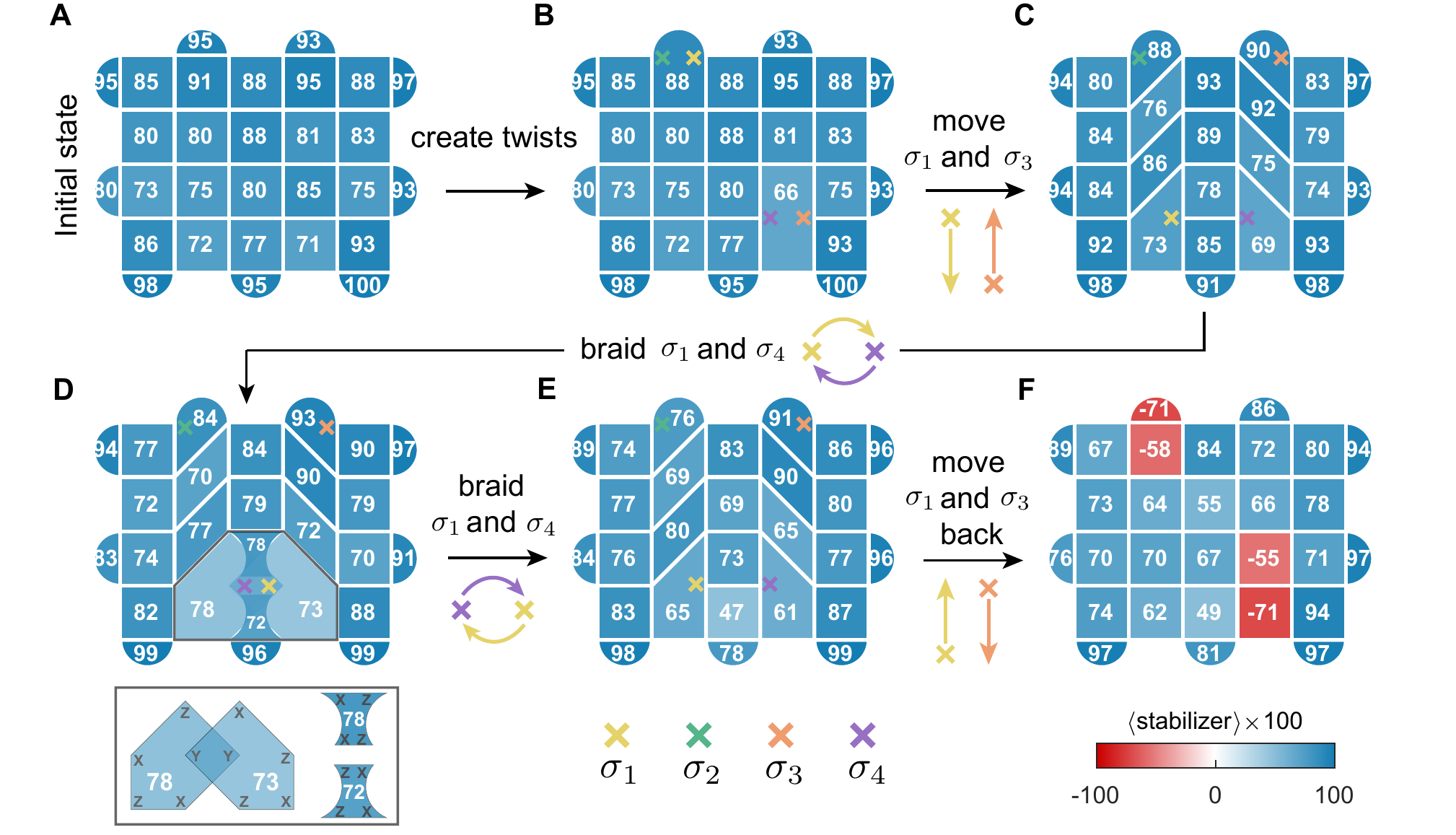}
	\caption{\label{fig-code-deformation-30} \textbf{Braiding twists through code deformation on processor II.} (\textbf{A})  The ground state of the surface code with no twist. (\textbf{B})  Two pairs of twists are created from the vacuum by deforming the stabilizer graph (the number of stabilizers is reduced by two). (\textbf{C})  Two pairs of twists are separated by code deformation, which in turn are implemented by a sequence of two-qubit gates. (\textbf{D})  Braiding ${\sigma _1}$  and ${\sigma _4}$ (${B_{14}}$). (\textbf{E})  Exchanging the positions of ${\sigma _1}$  and ${\sigma _4}$  again, such that all twists are in the same positions as in (\textbf{C}). The history of braiding twists can not be detected by local observables. (\textbf{F})  Moving ${\sigma _1}$ and ${\sigma _3}$ back and annihilating all twists. Two fermion charges are detected as the fusion results of the braided two pairs of twists, which are created from the vacuum in pairs originally.
	}
\end{figure}

\begin{figure}[htb]
	\centering\includegraphics[width=1\linewidth]{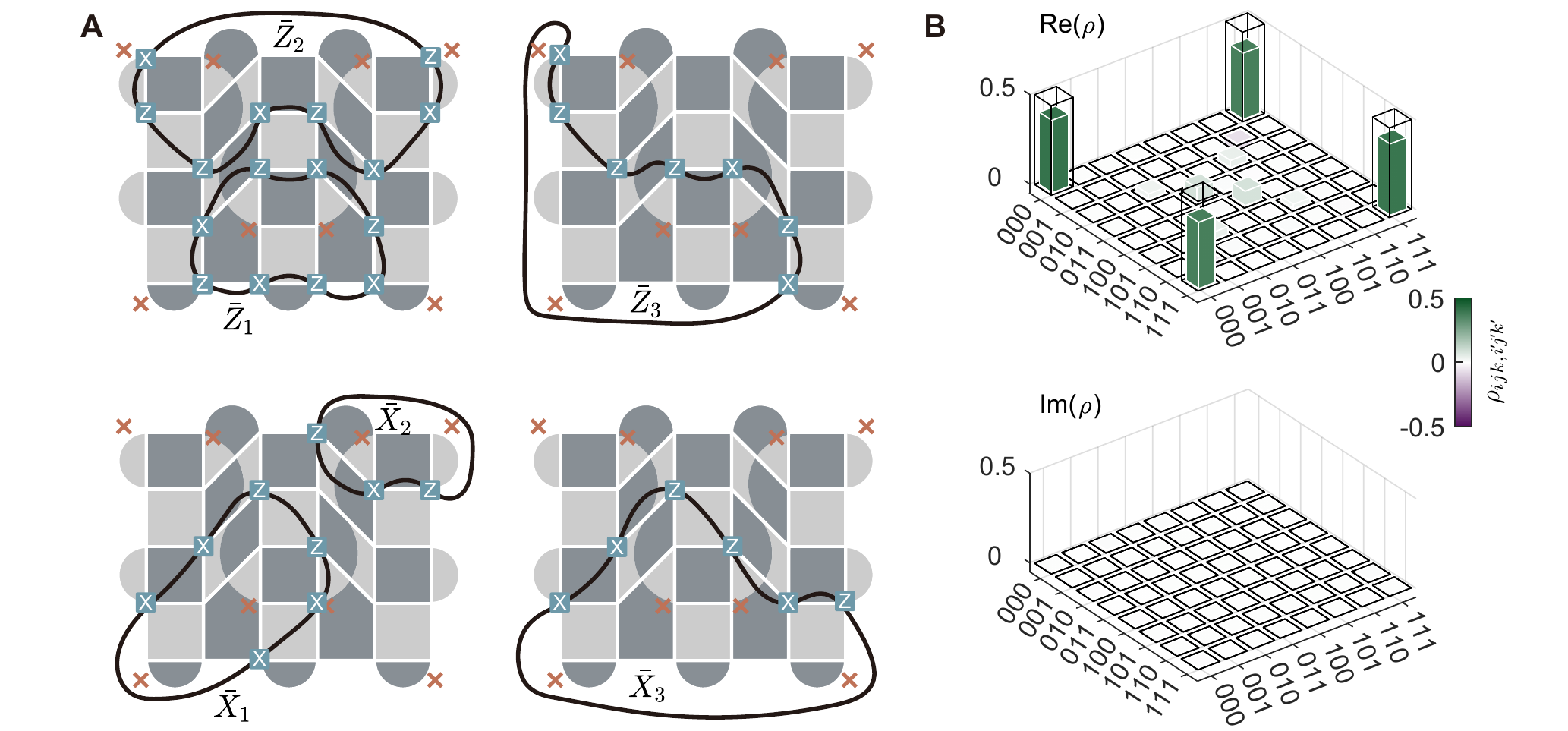}
	\caption{\label{fig-ghz-30}\textbf{Tomography of the logical GHZ state prepared on quantum processor II.} (\textbf{A})  The string operators corresponding to the logic ${\bar{Z}_1}$, ${\bar{Z}_2}$, ${\bar{Z}_3}$, ${\bar{X}_1}$, ${\bar{X}_2}$, and ${\bar{X}_3}$ on the three logical qubits encoded by the eight twists marked with the cross symbol ``$\times$". Using these logical operators, we perform  quantum  tomography of the logical state prepared in Fig. \ref{fig-code-deformation-30}\textbf{C}. The tomography is carried out by measuring the corresponding string operators on the physical state. %
		(\textbf{B})  The density matrix reconstructed from experimental measurement results of the prepared logical state. The fidelity between the prepared logical state and  the ideal three-qubit GHZ state ${1 \over {\sqrt 2 }}\left( {\left| {\overline {000} } \right\rangle  + \left| {\overline {111} } \right\rangle } \right)$ is $0.771 \pm 0.004$.}
\end{figure}

Here we present the results of simulating the non-Abelian statistics of twists with 30 qubits on quantum processor II using code deformation. The results are shown in Figs. \ref{fig-ground-state-30} to \ref{fig-code-deformation-30}. We also give the experimental result of quantum state tomography of logical GHZ state on 30 qubits in Fig. \ref{fig-ghz-30}, corresponding to Fig. 4 of the main text.

\end{document}